\documentclass
	[aps,pre,superscriptaddress, 
	preprint, showpacs,
	 amsmath, amssymb]{revtex4}

\usepackage{verbatim}

\usepackage{hyperref}
\usepackage{graphicx}
\usepackage{epsfig,amsmath,amsfonts,amssymb,wasysym,xcolor,hyperref}
\usepackage{soul}

\newcommand{\eref}[1]{Eq.~(\ref{#1})}
\newcommand{\abs}[1]{\ensuremath{\left\lvert #1 \right\rvert}}
\newcommand{\at}[2]{\ensuremath{\left.#1\right\rvert_{#2}}} 
\renewcommand{\c}[1][]{\ensuremath{c_{0#1}}} 
\newcommand{\cs}[1][]{\ensuremath{c_{#1}}} 
\newcommand{\ccf}[1][]{critical Casimir force#1}
\newcommand{\diff}[2]{\frac{\partial #1}{\partial #2}}
\newcommand{\ddiff}[2]{\frac{\partial^2 #1}{\partial {#2}^2}}
\newcommand{\fe}[1][]{\ensuremath{\mathfrak{f}_{#1}}}
\newcommand{\feV}[1][]{\ensuremath{\mathrm{f}_{#1}}}
\newcommand{\f}{\ensuremath{f_C}}
\newcommand{\fs}[1][]{\ensuremath{\vartheta_{#1}}}
\newcommand{\ccapp}{\ensuremath{\left\vert\Delta_{\of{+,+}}\right\vert}}
\newcommand{\fd}[1]{\ensuremath{\left[ #1 \right]}}

\newcommand{\Ham}[1][]{\ensuremath{\mathcal{H}_{#1}}}
	\newcommand{\h}[1][1]{\ensuremath{ h_{0,#1}}}
	\newcommand{\ho}[1][]{\ensuremath{ h_{0#1}}}
	\newcommand{\hs}[1][1]{\ensuremath{ h_{#1}}}
	
	\newcommand{\hxs}[2][1]{\ensuremath{\mathsf{h}_{#1}^{#2}}}
		
	\newcommand{\hb}[1][]{\ensuremath{h_{b#1}}}
\newcommand{\hot}[1]{\ensuremath{\mathcal{O}\left(#1\right)}}
\newcommand{\rscal}[1]{\ensuremath{{#1}^*}}
\newcommand{\sfct}[1][]{scaling function#1}
\newcommand{\sfcts}[1][]{scaling functions#1}
\newcommand{\fsf}[1][]{scaling function of the critical Casimir force#1}
\newcommand{\fsfs}[1][]{scaling functions of the critical Casimir force#1}
\newcommand{\of}[1]{\ensuremath{\left(#1\right)}}
\newcommand{\op}[1][]{order parameter#1}
\newcommand{\Op}[1][]{\ensuremath{\Phi_{#1}}}
\newcommand{\Ops}[1][]{\ensuremath{\mathcal M_{#1}}}
\newcommand{\opp}[1][]{order parameter profile#1}
\newcommand{\set}[1]{\ensuremath{\left\lbrace #1 \right\rbrace}}
\newcommand{\Sf}[1][]{\ensuremath{\mathfrak{S}_{#1}}}
\newcommand{\T}{\ensuremath{\mathcal T_{zz}}}
\newcommand{\tcb}[1][]{\ensuremath{T_{c,b}}}
\newcommand{\tcs}[1][]{\ensuremath{T_{c,s}}}
\newcommand{\tauch}[1][]{\ensuremath{\tau_{c#1}}}
\newcommand{\tauw}[1][]{\ensuremath{\tau_{w#1}}}
\renewcommand{\u}[1][]{\ensuremath{\mathfrak{u}_{#1}}}

\newcommand{\yc}[1][]{\ensuremath{y_{c#1}}}

\newcommand{\zs}[1][]{\ensuremath{\zeta_{#1}}}
	\newcommand{\dOp}[1][]{\ensuremath{\delta\Op{#1}}}
	\newcommand{\zp}{\ensuremath{z_{+}}}
	\newcommand{\zm}{\ensuremath{z_{-}}}

\begin{document}

\title{Crossover of Critical Casimir forces between different surface universality classes}
\date{\today}

\author{T. F. Mohry}
\email{mohry@mf.mpg.de}
\affiliation{Max-Planck-Institut f{\"u}r Metallforschung, Heisenbergstra{\ss}e 3, 70569 Stuttgart, Germany}
\affiliation{Universit{\"a}t Stuttgart, Institut f{\"u}r Theoretische und Angewandte Physik, Pfaffenwaldring 57, 70569 Stuttgart, Germany}

\author{A. Macio{\l}ek}
\affiliation{Max-Planck-Institut f{\"u}r Metallforschung, Heisenbergstra{\ss}e 3, 70569 Stuttgart, Germany}
\affiliation{Universit{\"a}t Stuttgart, Institut f{\"u}r Theoretische und Angewandte Physik, Pfaffenwaldring 57, 70569 Stuttgart, Germany}
\affiliation{Institute of Physical Chemistry, Polish Academy of Sciences, Kasprzaka 44/52, PL-01-224 Warsaw, Poland}

\author{S. Dietrich}
\affiliation{Max-Planck-Institut f{\"u}r Metallforschung, Heisenbergstra{\ss}e 3, 70569 Stuttgart, Germany}
\affiliation{Universit{\"a}t Stuttgart, Institut f{\"u}r Theoretische und Angewandte Physik, Pfaffenwaldring 57, 70569 Stuttgart, Germany}

\begin{abstract}
In confined systems near a continuous phase transition the long-ranged fluctuations of the corresponding order parameter are subject to boundary conditions. These constraints result in so-called critical Casimir forces acting as effective forces on the confining surfaces. For systems belonging to the Ising bulk universality class corresponding to a scalar order parameter the critical Casimir force is studied for the film geometry in the crossover regime characterized by different surface fields at the two surfaces. The scaling function of the critical Casimir force is calculated within mean field theory. Within our approach, the scaling functions of the critical Casimir force and of the order parameter profile for finite surface fields can be mapped by rescaling, except for a narrow crossover regime, onto the corresponding scaling function of the so-called normal fixed point of strong surface fields. 
In the crossover regime, the critical Casimir force as function of temperature exhibits more than one extremum and for certain ranges of surface field strengths it changes sign twice upon varying temperature.
Monte Carlo simulation data obtained for a three-dimensional Ising film show similar trends. The sign of the critical Casimir force  can be inferred from the comparison of the order parameter profiles in the film and in the semi-infinite geometry.
\end{abstract}


\maketitle

\section{Introduction \label{sec:intro}}

Finite-size contributions to the free energy of a fluid confined  between two planar surfaces, separated by a distance L, give rise to an effective force per unit area between the surfaces, or an excess pressure. This so-called solvation force $f_{solv}$ depends on $L$, the thermodynamic state of the bulk fluid, the fluid-fluid interactions, and the two substrates potentials \cite{evans}. At the bulk critical point of the fluid the solvation force acquires a universal contribution which is long-ranged in $L$. 
This fluctuation induced effective force is called the critical Casimir force \cite{FdG,krech99,revs1}.

The  critical  Casimir effect is a subject of considerable theoretical and experimental interest, involving experiments for wetting films near critical end points \cite{GC99,GC2002,FYP2005,RBM2007} and for colloidal systems in the presence of a critical solvent \cite{nature, Gambassi-et:2009, Soyka-et:2008, Troendle-et:2009}. 
The sensitive temperature dependence  of the \ccf{} can be exploited in the latter systems in order to control the collective behavior  of colloidal particles, such as their aggregation behavior, which opens up application perspectives in many areas of material science. 
According to the accumulated knowledge, the sign of the \ccf{}  can be selected by suitable surface treatments \cite{krech99,BTD200book, Gambassi:2009}. Recently, a continuous tuning has been achieved experimentally for a colloidal particle in a critical solvent and near a substrate with a gradient in its preferential adsorption properties for the two species forming the binary liquid mixture as a solvent \cite{nhb}. It is very encouraging that the adsorption preference of a substrate can be changed continuously between strong adsorption of one species to strong adsorption of the other species of a binary liquid mixture by tuning the chemical composition of a monomolecular overlayer only, without altering the bulk material of the confining substrates. As will be discussed later, in the present context this amounts to continuously tune a surface field $h_1$, which expresses this preference and breaks the symmetry of the \op[,] between $+\infty$ and $-\infty$.

Such a tunability of critical Casimir forces towards repulsion might be relevant for micro- and nano-electromechanical systems in order to prevent stiction due to the omnipresent attractive 
quantum mechanical Casimir forces \cite{casimir48} - for exceptions see Ref.~\cite{Capasso}. 
In order to achieve repulsive quantum Casimir forces, rather complex systems have been considered but they are not yet experimentally established \cite{Rodriguez}.
Here we show theoretically that, for suitably prepared system parameters,  switching the sign of the critical Casimir force can be achieved not only by varying the surface fields but also via minute temperature changes. This occurs if the adsorption properties of the confining surfaces differ significantly, for example, if each surface attracts a different component of the binary liquid mixture but one does so weakly and the other strongly.

To be specific, we investigate theoretically the properties of the  critical Casimir force in thin films of systems  belonging to the Ising universality class (UC) focusing on the crossovers between various surface UCs \cite{Binder1983,Diehl1986}, i.e., systems for which one or both surfaces give rise to relatively weak adsorption. Representatives of this class are simple fluids, binary liquid mixtures, or Ising ferromagnets.  The temperature dependence of the critical Casimir force, its sign, and its strength depend on the nature of the confining surfaces, which impose specific boundary conditions (BCs) on the relevant \opp[.] In this context, so far only the cases of strongly adsorbing or neutral surfaces forming various surface UCs \cite{krech99,BTD200book, Gambassi:2009} have been studied theoretically, for both symmetric and antisymmetric BCs. In order to understand and thus to be able to control the aforementioned tunability of the critical Casimir forces, here we focus on the crossovers between these different surface  UCs, i.e., systems for which one or both surfaces give rise to relatively weak adsorption.

In order to calculate the critical Casimir force in the spirit of fieldtheoretical renormalization group theory we use the Landau-Ginzburg model in the film geometry.
Within this approach the surfaces $1$ and $2$ are characterized by surface fields \h[i], $i=1,2$, conjugated to the \op{} at the surface, and by so-called surface enhancement parameters \c[,i], $i=1,2$, describing the tendency of the system to order at the surface \cite{Binder1983,Diehl1986}. Our results have been obtained numerically within  mean field theory (MFT) as the lowest order contribution in a systematic $4-d$ expansion in $d$ spatial dimensions. The universal scaling functions of the order parameter profile and of the critical Casimir force have been calculated and thoroughly analyzed for the crossover between the so-called normal and the so-called special surface transition and for the crossover between the normal and the ordinary surface transition \cite{Binder1983,Diehl1986}, i.e., for various values of the parameters \h[i] and \c[,i]. It turns out that depending on the choice of these surface properties,  the critical Casimir force can  change its sign once or even twice  upon varying the temperature. We also  propose a simple criterion relating the sign of the critical Casimir force to the values of the order parameter at the surfaces and in the bulk.

Recently, the crossover behavior in the same type of model, but for symmetry-preserving BCs only (which enables one technically to go beyond MFT) has been studied by field-theoretic methods \cite{SD2008}.
Explicit two-loop renormalization group calculations show that the critical Casimir force can  be of either sign depending on the surface enhancement parameters \c[,1] and \c[,2]. However, these results are not applicable for fluid systems, because generically these are exposed to symmetry breaking surface fields. Here we study the experimentally relevant case of tuning surface fields.
In the presence of arbitrary surface fields both Monte Carlo (MC) simulation data \cite{Vasilyev-et:2010} for the three-dimensional Ising model in the film geometry as well as exact results for two-dimensional Ising strips \cite{Abraham-et:2009,Nowakowski-et:2008,Nowakowski-et:2009} show similar trends in the behavior of the critical Casimir force.

Our presentation is organized as follows: In Sec.~\ref{sec:model} we introduce the model and briefly present the relevant basic theoretical facts concerning finite-size scaling and surface UCs. In Sec.~\ref{sec:res} we report results of our calculations for the crossover between the special and the normal transition (Subsec.~\ref{ssc:spnorm}) and the crossover between the ordinary and the normal transition (Subsec.~\ref{ssc:ordnorm}). In Sec.~\ref{sec:dim}  our results pertinent to four spatial dimensions are compared to results for $d=2$ available in literature  \cite{Nowakowski-et:2008,Nowakowski-et:2009, Abraham-et:2009}
and we provide a comparison with MC simulation data for $d=3$ Ising films \cite{Vasilyev-et:2010}. Section~\ref{sec:summary} summarizes our results.

\section{Theoretical background  \label{sec:model}}

\subsection{The model\label{ssc:lgh}}
In order to calculate the \opp{} and the \ccf{} we use the standard reduced Landau-Ginzburg Hamiltonian $\overline{\Ham}=\widetilde{\Ham}/\of{k_B T}$ (in units of the thermal energy $k_B T$ and thus dimensionless) describing a system with $O(N)$ symmetry. For the film geometry with planar, laterally homogeneous surfaces and within MFT the \op{} (OP) profile \Op{} depends only on the spatial variable $z$ orthogonal to the surfaces so that $\overline{\Ham}=A\Ham$ with
\begin{equation}
\label{eq:slitham}
\begin{split}
\Ham \fd{\Op\of{z}}=
	&\int_{-L/2}^{L/2} \mbox{d}z\; 
		\set{
		\frac{1}{2} \of{\diff{\Op\of{z}}{z} }^2
		+ \frac{\tau}{2} \Op^2 \of{z}
		+ \frac{g}{4!} \Op^4 \of{z}
		- h_{0,b} \Op\of{z}
		} \\
	& + \frac{\c[,1]}{2} \Op[1]^2 - \h[1] \Op[1]
	+ \frac{\c[,2]}{2} \Op[2]^2 - \h[2] \Op[2]
\end{split}
\end{equation}
and where $A$ is the macroscopically large, \of{d-1}-dimensional area of one of the equally sized confining surfaces.
Corresponding to the Ising UC studied here \Op{} is a scalar. $\Op[1]\equiv \Op\of{z=-L/2}$ and $\Op[2]\equiv \Op\of{z=L/2}$ are the values of the \op{} at the confining walls. The coefficient $\tau\propto \frac{T-\tcb}{\tcb}$, where $\tcb$ is the bulk critical temperature, changes sign at bulk criticality. The coupling constant $g>0$ stabilizes \Ham{} for $T<\tcb$. In the following we assume that the ordering bulk external field \hb[,0] is zero, i.e., we focus on the critical concentration of the fluid. The effects of the surfaces on the system are captured by the surface fields \h[i] and by the surface enhancements \c[,i] as will be discussed in Subsec.~\ref{ssc:surfuni}. In the sense of renormalization group theory Eq.~\eqref{eq:slitham} captures all relevant scaling fields and thus is able to predict the leading universal behavior of critical films \cite{Binder1983,Diehl1986}.

In the film geometry the \ccf{} per area $A$ of one of the equally sized confining surfaces and in units of $k_B T$ is given by
\begin{equation}
\label{eq:defccf}
	\f \equiv -\diff{\fe^{ex}}{L}, 
\end{equation}
where the excess free energy per area and in units of $k_B T$ is defined as 
\begin{equation}
\label{eq:defccfex}
	\fe^{ex}\equiv \of{\feV - \feV[b]}L/\of{k_B T}.
\end{equation}
Here $\feV$ is the singular contribution to the total free energy of the film per volume $V=LA$ and $\feV[b]$ is the singular part of the bulk free energy density.

Within MFT the analysis of the \ccf{} leads to the Euler-Lagrange equation (ELE)
\begin{equation}
\label{eq:ele}
	\ddiff{\Op}{z} = \tau \Op + \frac{g}{6} \Op^3
\end{equation}
with the BCs
\begin{align}
	\at{\diff{\Op}{z}}{z=-L/2} &= \c[,1] \Op[1] -\h[1]
			\label{eq:bc1}
	\intertext{and}
	\at{\diff{\Op}{z}}{z=L/2} &= -\of{\c[,2] \Op[2] -\h[2]}.
			\label{eq:bc2}
\end{align}
Instead of using the definition of the \ccf{}  given by Eq.~\eqref{eq:defccf}, it is more convenient to use the thermal average of the  \of{z,z}-component of the stress tensor \cite{Eisenriegler-et:1994, krech99} (we omit here the  brackets $\langle \cdot \rangle $ indicating the thermal average):
\begin{equation}
\label{eq:stresstensor}
\T \fd{\Op}
	=\frac{1}{2}\of{\diff{\Op}{z}}^2
	-\frac{\tau}{2}{\Op}^2\of{z}
	-\frac{g}{4!}{\Op}^4\of{z},
\end{equation}
which is spatially constant throughout the film including the surfaces. In order to obtain the \ccf{}  the bulk contribution has to be substracted:
\begin{equation}
\label{eq:ccf_via_stress}
\f = \T \fd{{\Op}} -\T \fd{{\Op[b]}},
\end{equation}
where the bulk value \Op[b] is
\begin{equation}
\label{eq:opb}
\Op[b]= \pm\Theta\of{-\tau}\sqrt{\frac{-6\tau}{g}}
\end{equation}
and where $\Theta$ is the Heaviside step function.

Dimensional analysis leads to the following scaled, dimensionless quantities:
\begin{align}
	\zs &= z/L & 
		\label{eq:zeta} 
		\\
	\Ops\of{\zs} &= \sqrt{\frac{g}{6}} L^{\beta/\nu} \Op\of{\zs L}&
		\of{\beta/\nu}_{MFT} &= 1
		\label{eq:ops}
		\\
	y &= \tau L^{1/\nu} &
		\nu_{MFT} &= 1/2
		\label{eq:y}
		\\
	\cs[i]& = \c[,i] L^{\phi/\nu} &
		\of{\phi/\nu}_{MFT} &= 1
		\label{eq:cs}
		\\
	\hs[i]& = \sqrt{\frac{g}{6}}\h[i] L^{\Delta_1^{sp}/\nu} &
		\of{\Delta_1^{sp}/\nu}_{MFT} &= 2,
		\label{eq:hs}
\end{align}
where $\phi$ is the crossover exponent and  $\Delta_1^{sp}$ is the surface counterpart of the bulk gap exponent $\Delta$. In terms of these scaled variables
the Hamiltonian in \eref{eq:slitham} takes the scaled form
\begin{equation}
\label{eq:hamscaled}
\begin{split}
\Ham \fd{ \Op\of{z} } = &
	\frac{6}{g} L^{-3}
	\int_{-1/2}^{1/2}\mbox{d}{\zeta} \set{ 
		\frac{1}{2} \of{\diff{\Ops\of{\zeta}}{\zeta} }^2
		+ \frac{y}{2} \Ops^2\of{\zeta} 
		+ \frac{1}{4} \Ops^4\of{\zeta} }\\
	& + \frac{\cs[1]}{2} \Ops[1]^2 - \hs[1] \Ops[1]
	+ \frac{\cs[2]}{2} \Ops[2]^2 - \hs[2] \Ops[2],
\end{split}
\end{equation}
where within MFT the prefactor $6/g$ is undetermined. 
In these scaled units the bulk limit (which minimizes the integral in \eref{eq:hamscaled} for constant $\Ops{}=\Ops[b]$) and the correlation length, defined via the exponential decay of the two-point correlation function for $T\neq\tcb$, are
\begin{equation}
\label{eq:scaledopb}
	\Ops[b] = \pm\Theta\of{-y} \abs{y}^{1/2}
\end{equation}
and
\begin{equation}
\label{eq:slit_scaledxi}
	\xi/L = 
		\begin{cases}
		y^{-1/2} & y>0 \\
		\of{-2y}^{-1/2} & y<0
		\end{cases},
\end{equation}
respectively.

The above MFT considerations can be put into the general context of scaling theory. In accordance with this the \opp{} in a film exhibits the following scaling behavior \cite{Binder1983}:
\begin{equation}
\label{eq:opp_scaling_1}
	\Op\of{z,t,L}=\Op[b]\of{t}{\cal \widetilde{M}}\of{z/\xi,L/\xi},
\end{equation}
where $\xi\of{t=\of{T-\tcb}/\tcb\to\pm 0}=\xi_0^{\pm}\abs{t}^{-\nu}$; $\xi_0^{\pm}$ are nonuniversal amplitudes with a universal ratio $\of{\xi_0^{+}/\xi_0^{-}}_{MFT}=\sqrt{2}$. $\Op[b]\of{t}=\Op[b,0]\abs{t}^{\beta}$ is the bulk \op{} (we take $\Op[b]\of{t>0}$ to be $\Op[b]\of{-t}$ valid for $T<\tcb$) with the (only) second nonuniversal amplitude \Op[b,0]. In \eref{eq:opp_scaling_1} ${\cal\widetilde{M}}\of{x,\tilde{y}}$ is a universal \sfct{} which is normalized such that ${\cal\widetilde{M}}\of{x\to\infty,\tilde{y}=\infty}=\Theta\of{-t}$. The \sfct{} ${\cal \widetilde{M}}\of{x,\tilde{y}}$ is suitable for capturing the semi-infinite limit. For finite film thickness it is helpful to rewrite \eref{eq:opp_scaling_1} in terms of a different, also universal scaling function
$\Ops\of{\zs=z/L,y=\of{\text{sign t}}\of{L/\xi}^{1/\nu}=t\of{L/\xi_0^+}^{1/\nu}}$:
\begin{equation}
\label{eq:opp_scaling_2}
	\Op\of{z,t,L}=
		\Op[b]\of{t}\of{\xi/L}^{\beta/\nu}
			\Ops\of{z/L,t\of{L/\xi^+_0}^{1/\nu}}
		=\Op[b,0]\of{L/\xi^+_0}^{-\beta/\nu}
			\Ops\of{\zs,y}.
\end{equation}
Note that in \eref{eq:opp_scaling_2} we have used as expression for $\xi$ the one above $T_c$ also for $t<0$ so that here $\xi(t)=\xi^+_0|t|^{-\nu}$ for all $t$; 
this yields a scaling variable $y$ which is analytic in $t$. The universal \sfct{} $\Ops\of{\zs,y}$ has the property and is normalized such that, for $\zs \abs{y}^{\nu}=const$, 
$\Ops\of{\zs\to 0,y\to\infty}=\Theta\of{-y}\abs{y}^{\beta}\equiv\Ops[b]\of{y}$
which implies $\lim\limits_{\stackrel{L\to\infty,}{t\text{ fixed}}}\fd{ \of{\xi/L}^{\beta/\nu}\Ops\of{z/L,\of{\text{sign }t} \of{L/\xi}^{1/\nu}}}=\Theta\of{-t}$.
In the following $\Ops[b]\of{y}$ will be also called as bulk \op{}.
The above MFT expressions (Eqs.~\eqref{eq:zeta}-\eqref{eq:hs}) are in line with these general properties by noting that within MFT one has $\tau=\of{\xi_0^+}^{-2}t$.

Finite-size scaling theory \cite{Binder1983,Diehl1986,Barber1983, Privman:1990} and renormalization group theory \cite{Krech-et:1992} for the film geometry predict that the \ccf{}  takes on the following scaling form:
\begin{equation}
\f\of{\tau,\c[,1],\h[1],\c[,2],\h[2]; L}
	= \of{d-1} L^{-d}
		\fs \of {y, \cs[1] ,\hs[1],\cs[2],\hs[2]}.
\label{eq:ccfscaling}
\end{equation}
Within MFT the universal \sfct{} \fs{} of the \ccf{} can be determined only up to an undetermined prefactor $\sim1/g$ (compare the note after Eq.~\eqref{eq:hamscaled}). An appropriate way to cope with this is to express \fs{} in units of the critical Casimir amplitude $\Delta_{\of{+,+}}$ at \tcb{} for fixed point BCs $\hs[1]=\hs[2]=\infty$, which carries the same undetermined prefactor. Accordingly, all our MFT results expressed in units of $\Delta_{\of{+,+}}\of{d=4}=-\of{6/g}\frac{4}{3}K^4 <0$, where $K\equiv K\of{1/2}$ is the complete elliptic integral of the first kind \cite{Krech1997}, are independent of this undetermined prefactor and are therefore accessible to comparisons with results obtained by other theoretical techniques or experimentally.
(Note that within MFT, i.e., $d=4$, $g$ is dimensionless.)

\subsection{Surface universality classes \label{ssc:surfuni}}

Near a surface the system properties differ from their bulk values. For example, the tendency to order is reduced due to missing neighbors and the coupling between the ordering degrees of freedom at the surface can differ from its bulk value. In \eref{eq:slitham} these surface effects are captured and  characterized by the surface enhancements \c[,i], which preserve the $O(N)$ symmetry. In addition, the surface can favor one bulk phase over the other. In Eq.~\eqref{eq:slitham} this explicit breaking of the symmetry is described by the surface fields \h[i]. For detailed discussions of surface criticality see Refs.~\cite{Binder1983,Diehl1986,diehl1997} and  Ref.~\cite{Kuipers-et:2009}. In the latter, microscopic expressions for $h_0$  and $c_0$  are derived by using density functional theory. 

A semi-infinite system the surface of which  has a reduced tendency to order belongs to the so-called ordinary surface UC (in the following labeled by $\u=ord$). The corresponding fixed point values of the surface parameters are $\c^{ord}=\infty$ and $\ho^{ord}=0$. 
Because \c{} is a so-called dangerous irrelevant variable, it may not simply be set to its fixed point value \cite{Diehl1986}, but  rather the following linear scaling field has to be considered:
\begin{equation}
\label{eq:ordscalfield}
\hxs[0]{ord}=\ho/\c^a,
\end{equation}
with the scaling exponent
\begin{equation}
\label{eq:ordexp}
a= \left. \frac{\Delta_1^{sp} - \Delta_1^{ord}}{\phi}\right\vert_{MFT} = 1.
\end{equation}
Thus the dependences on the scaling variables in Eq.~\eqref{eq:ccfscaling} reduce to the dependence on a single scaling variable per surface, which, within MFT, is $\hxs[i]{ord}=\hs[i]/\cs[i]\sim \of{\h[i]/\c[,i]} L^{\Delta_1^{ord}/\nu}$ where $\Delta_1^{ord}=1/2$ within MFT.

Systems with surfaces, which do not break explicitly the $O(N)$ symmetry of the \op{} but locally enhance the ordering, belong to the so-called extraordinary surface UC characterized by the fixed point values $\c^{ex}=-\infty$ and $\ho^{ex}=0$. Such surfaces order at a surface transition temperature $\tcs>\tcb$. 
In the parameter space spanned by \of{T,\c,\ho=0} the transition lines of the ordinary transition, the extraordinary transition, and the surface transition meet at the multicritical point \of{\tcb,\c^*} of the so-called special transition (in the following labeled by $\u=sp$). Within MFT the value of the surface enhancement at the special transition is $\c^*=0$. Near the special transition both \ho{} and \c{} are relevant.

Surfaces which prefer one of the bulk phases over the other  break the symmetry of the \op{} explicitly. Semi-infinite systems with such surfaces belong to the so-called normal UC (in the following labeled by $\u=\pm$, depending on the sign of the surface field) characterized by the fixed point $\ho^{norm}=\infty$. For both  normal and  extraordinary UCs the  surface is ordered even above \tcb. Explicit calculations for the Ising model \cite{Burkhardt-et:1994} and renormalization group analyses \cite{diehl1997} reveal the equivalence of the normal and the extraordinary transitions.

For films the character of both surfaces matter. Therefore  the UCs for the systems in the film geometry can be labeled by the two UCs of the corresponding surfaces in the semi-infinite geometry. This interpretation is meant implicitly when in the following the crossover behavior in the film geometry is  named after  the crossover between  surface UCs in the corresponding semi-infinite systems.

At the fixed points and within MFT the scaling functions $\fs[{\u[1],\u[2]}]$, $\u[i]\in\set{\pm,sp,ord}$, are known analytically \cite{Krech1997,Maciolek-et:2007}. In order to obtain the {\sfcts} for the surface parameters \c[,i] and \h[i] being off their fixed point values, we have minimized the Hamiltonian in Eq.~\eqref{eq:slitham} numerically.

\section{Results \label{sec:res}}

\subsection{Crossover from the special to the normal transition \label{ssc:spnorm}}

In this subsection we study the behavior of the {\sfcts} for various applied surface fields $h_{0,1}$ and $h_{0,2}$, including $h_{0,1}\ne h_{0,2}$, at vanishing surface enhancement parameters $c_{0,1}=c_{0,2}=0$.

\subsubsection{Rescaling \label{sss:rescale}}

For a wide range of finite values of the surface fields \hs[i], the variation of the \sfct{} $\fs\of{y}$ of the \ccf{} as a function of the scaled temperature $y$  (see \eref{eq:y}) is  very similar to the one corresponding to the fixed point solutions $\fs[\of{+,+}]\of{y}$  or $\fs[\of{+,-}]\of{y}$, depending on the sign of $\hs[1]\hs[2]$  (see Fig.~\ref{fig:spnorm_crossover}). As described below
this similarity can be specified quantitatively by a suitable rescaling 
of the scaling functions.

This rescaling idea is borne out by our observation that the OP profile  $\Op\of{z;\tau,\h[1],\h[2],L}$ of  the film of width $L$ exposed to {\it finite} surface fields  $h_{0,i}$, $i=1,2$, can be expressed in terms of the OP profile  $\Op\of{z;\tau,\h[1]=\infty,\h[2]=\pm\infty,\rscal{L}}\equiv\Op[\of{+,\pm}]\of{z;\tau,\rscal{L}}$ of a film of width $L^*>L$ with {\it infinite} surface fields, in such a way that the former profile is a portion of the latter one.
(We recall that for films with infinite surface fields the OP diverges at the surfaces.)
Since for both scaling functions of the OP profile the spatial scaling variable has the range  $-0.5\le \zeta \le 0.5$,  the scaling function corresponding to the film with finite surface fields has to be shifted and "stretched" in order to fit into the corresponding interval. The corresponding scaling variable and the amplitude of the \sfct{} 
may be different from those of $\Ops[\of{+,\pm}]\of{\zs;y}$. Therefore we make the ansatz
\begin{equation}
\label{eq:scalansatz}
	\Ops\of{\zeta;y,\hs[1],\hs[2]}=f\Ops[\of{+,\pm}]\of{r\of{\zeta-\zeta_0};\rscal{y}}.
\end{equation}
Comparing the ELE for both scaling functions, i.e., the scaled form of \eref{eq:ele},
$\Ops^{\prime\prime}=y\Ops+\Ops^{3}$ and $\Ops[\of{+,\pm}]^{\prime\prime}=y^*\Ops[\of{+,\pm}]+\Ops[\of{+,\pm}]^{3}$, where 
$\Ops^{\prime}=\partial\Ops/\partial\zs$,
we find \cite{note_rescl_genExponents}
\begin{equation}
\label{eq:rela}
f=r, \qquad y^*=r^{-2}y.
\end{equation}
The rescaling function $r=r\of{y,\hs[1],\hs[2]}$ and the shift $\zs[0]=\zs[0]\of{y,\hs[1],\hs[2]}$ are obtained from the scaled form of the BCs in Eqs.~\eqref{eq:bc1} and \eqref{eq:bc2}, $-\hs[1]=\Ops^{\prime}\of{-0.5;y,\hs[1],\hs[2]}$ and $\hs[2]=\Ops^{\prime}\of{0.5;y,\hs[1],\hs[2]}$, respectively:
\begin{align}
	r^2\Ops[\of{+,\pm}]^\prime\of{r \of{-0.5-\zs[0]};r^{-2}y}&=-\hs[1]&
	\label{eq:def_rs1}\\
	r^{2}\Ops[\of{+,\pm}]^\prime\of{r \of{0.5-\zs[0]};r^{-2}y}&=\hs[2].&
	\label{eq:def_rs2}
\end{align}
We note that $\zs[0]\neq0$ only if $\abs{\hs[1]}\neq\abs{\hs[2]}$.
With this the corresponding mapping for the \fsf{} can be read off from the expression for $f_C$ in terms of the stress-tensor, i.e., from Eqs.~\eqref{eq:stresstensor} and \eqref{eq:ccf_via_stress} by using Eqs.~\eqref{eq:ops}, \eqref{eq:scaledopb},  and \eqref{eq:ccfscaling}
(note that Eqs.~\eqref{eq:stresstensor} and \eqref{eq:ccf_via_stress} are valid only within MFT and thus require $d=4$ so that $g$ is dimensionless): 
\begin{equation}
\label{eq:rescCas}
\fs\of{y;h_1,h_2}=
	\of{d-1}^{-1}\frac{6}{g}
	\of{\frac{1}{2}\of{\Ops'}^2
		-\frac{y}{2}\of{{\Ops}^2-{\Ops[b]}^2}
		-\frac{1}{4}\of{{\Ops}^4-{\Ops[b]}^4}
	},
\end{equation}
from wich we find
\begin{equation}
\label{eq:rescCas1}
\fs\of{y;\hs[1],\hs[2]}=r^4\fs[\of{+,\pm}]\of{r^{-2}y}.
\end{equation}

The ranges of values of the surface fields \hs[i], for which the proposed mapping can be applied, is limited by the necessity to fulfill Eqs.~\eqref{eq:def_rs1} and \eqref{eq:def_rs2}.
For the \of{+,+} BC  $\Ops[\of{+,+}]^\prime\of{\zs;y}$ varies from $-\infty$ to 0 for $-0.5\leq\zs\leq0$ and from 0 to $\infty$  for $0\leq\zs\leq0.5$. Thus, for any  $\hs[1]>0$ and $\hs[2]\geq 0$ Eqs.~\eqref{eq:def_rs1} and \eqref{eq:def_rs2} have a solution for $r$ and $\zs[0]$.
We have checked the proposed mappings (Eqs.~\eqref{eq:scalansatz} and \eqref{eq:rescCas1}) by comparing them with the results of the numerical minimization of the Hamiltonian in \eref{eq:slitham} and we found excellent agreement. Examples for the mapping of \fs{} are shown in Fig.~\ref{fig:rescaled_force}.

In the case of opposing surface fields, $-\hs[1]\leq \hs[2]<0$ (here $\abs{\hs[1]}\geq\abs{\hs[2]}$, otherwise $1$ and $2$ have to be interchanged), the mapping is restricted by the upper bound for the slope $\Ops[\of{+,-}]^\prime\of{\zs;y}<0$. 
For fixed (but arbitrary) $\hs[1]>0$,  only films with
$-\hs[1]\leq\hs[2]\leq\hs[2,min]<0$ \cite{note_h2min} can be mapped onto films with \of{+,-} fixed point BC, where \hs[2,min] depends on both \hs[1] and $y$. For $y\geq0$, $\hs[2,min]\of{\hs[1],y}$ is obtained by evaluating $\Ops[\of{+,-}]^\prime$ in \eref{eq:def_rs2} at the position where \abs{\Ops[\of{+,-}]^\prime} is minimal, i.e., at $\zs[0]=0.5$ and we find
\begin{equation}
\label{eq:h2min}
	\hs[2,min]\of{\hs[1],y}= r_{min}^{2}{\cal M}_{(+,-)}^\prime\of{0;r_{min}^{-2}y}
\end{equation}
with $r_{min}$  given implicitly by \eref{eq:def_rs1}: 
\begin{equation}
\label{eq:rmin}
	r_{min}^{2}{\cal M}_{(+,-)}^\prime\of{-r_{min};r_{min}^{-2}y}=-\hs[1].
\end{equation}
Since $\Ops[\of{+,-}]^\prime\of{\zs}$ is defined for $-0.5\leq\zs\leq0.5$, \eref{eq:rmin} together with \eref{eq:h2min} yield the condition $0< r_{min}\leq 0.5$; 
$r_{min}$ attains its limiting value $0.5$ for $\hs[1]=\infty$ independent of $y\geq0$. \abs{\hs[2,min]} is largest for $y=0$ and decreases upon increasing $y$. For $y<0$ the determination of \hs[2,min] is more involved, because $\Ops[\of{+,-}]^\prime$ is no longer monotonic for $-0.5\leq\zs\leq0$.
For fixed \hs[1], the maximum of $\fs\of{y;\hs[1]>0,\hs[2]\leq\hs[2,min]\of{\hs[1],y=0}}$ as a function of $y$ moves towards $y=0$ upon  decreasing \hs[2] and we observe that for $\hs[2]=\hs[2,min]\of{\hs[1],y=0}$ the maximum occurs at \tcb.

It turns out, that for sufficiently strong surface fields the rescaling functions $r\of{y,\hs[1],\hs[2]}$ and $\zs[0]\of{y,\hs[1],\hs[2]}$ are well approximated by resorting to the \textit{s}hort \textit{d}istance behavior of \Ops{} and they become rescaling parameters which are independent of $y$.
One obtains (see Appendix~\ref{sec:app_rscal-sd})
\begin{equation}
	r_{sd} = \fd{1+l\of{\hs[1]}+l\of{\hs[2]}}^{-1}
	\label{eq:r_sd}
\end{equation}
and
\begin{equation}
	\zs[0]^{sd} = \of{l\of{\hs[2]}-l\of{\hs[1]}}/2,
	\label{eq:zs0_sd}
\end{equation}
where $l\of{\hs[i]}=2^{1/4}\abs{\hs[i]}^{-1/2}$.
The comparison with our numerical data shows, that for  systems with $\hs[1],\hs[2]\gtrsim10$ this approximation works very well for the whole range of $y$: for these films the difference between using $r\of{y;\hs[1],\hs[2]}$ and $r_{sd}\of{\hs[1],\hs[2]}$ for the mapping is not visible on the scale of Fig.~\ref{fig:rescaled_force}.
On the other hand, for values $\hs[1],-\hs[2]\gtrsim10$ 
this approximation is only valid for $y>y_{max}$, where $y_{max}$ is the position of the maximum of \fs{}.
As expected, our numerical data show, that at \tcb{} for those values of the surfaces fields for which the short-distance approximation holds, the OP profile of the film near the surfaces can be well aproximated by the corresponding OP profiles of the semi-infinite sytem. For weaker surface fields the profiles differ significantly - even at the surfaces.

The deviation of the \sfct{} $\fs\of{y;\hs[1],\hs[2]}$ of the \ccf{} from the fixed point \sfct{} $\fs[\of{+,\pm}]\of{y}$ in leading order of \hs[1] and \hs[2] is obtained from the mapping given by \eref{eq:rescCas1} and from $r_{sd}$ in \eref{eq:r_sd}:
%
\begin{multline}
\label{eq:fsfapproachfix}
\fs\of{y;\hs[1],\hs[2]}-\fs[\of{+,\pm}]\of{y} 
	\simeq \\ 
	- 2^{9/4} \of{\abs{\hs[1]}^{-1/2}+\abs{\hs[2]}^{-1/2}}
	\set{\fs[\of{+,\pm}]\of{y}-\frac{y}{2}\fs[\of{+,\pm}]^{\prime}\of{y}},
	\quad \abs{\hs[1]},\abs{\hs[2]}\to \infty.
\end{multline}
%
Our numerical data for the scaling functions are in agreement with \eref{eq:fsfapproachfix} for $\abs{\hs[1]},\abs{\hs[2]}\gtrsim 1000$. This algebraic behavior explains the slow convergence of the \sfct{} \fs{} towards the fixed point \sfct{} \fs[\of{+,\pm}] as apparent from Fig.~\ref{fig:spnorm_crossover}. We note that, since the term in curly brackets in \eref{eq:fsfapproachfix} is comparable with $\fs[\of{+,\pm}]$ itself, the relative deviation for, e.g., $\hs[1]=1000$ and $\hs[2]=\infty$ is still about $15\%$. 
Our data for the critical Casimir amplitude $\Delta_+\of{\hs[1]} \equiv\Delta\of{\hs[1],\hs[2]=\hs[1]} \equiv\fs\of{y=0;\hs[1],\hs[2]=\hs[1]}$ shown in  Fig.~\ref{fig:spnorm_casampl-various}(a) display an algebraic behavior which is in accordance with \eref{eq:fsfapproachfix}.

\subsubsection{Weak surface fields \label{ssc:smallsurffields}}

For weak surface fields $\h[1]=\h[2]$, i.e., if the length scale $l_{sp}=\of{\sqrt{g/6}\h[1]}^{-\nu/\Delta_{1}^{sp}}$ \cite{CA-83} associated with the surface field in the semi-infinite geometry dominates over $L$, 
both surfaces approach the special transition (we recall that here we consider the case $\cs[1]=\cs[2]=0$) and the \opp[s] do not vary substantially across the film (data not shown). In this case the square gradient term in \Ham{} (\eref{eq:slitham}) can be neglected and the \op{} can be approximated to be constant so that the unscaled free energy (per $k_b T$ and $A$) is
\begin{equation}
\label{eq:fesmallh}
	\fe\of{\h[1]=\h[2]\to 0}
	= L \of{\frac{\tau}{2}\Op[{\h[1]}]^2+\frac{g}{4!}\Op[{\h[1]}]^4}
	- 2\, \h[1] \,\Op[{\h[1]}].
\end{equation}
Comparison with \eref{eq:slitham} shows, that the surface fields $\h[1]=\h[2]\to 0$ act like an effective bulk field $\hb[,0]=2\h[1]/L$. 
This is in line with the analysis by Nakanishi and Fisher \cite{Nakanishi-et1983} of the shift  of the critical point and of the phase boundary in films, which shows  that for $\h[1]=\h[2]>0$ the phase boundary is shifted towards negative values of the bulk ordering field \h[b].
Minimization of the free energy in \eref{eq:fesmallh} with respect to \Op[{\h[1]}] yields at $\tau=0$ the critical value of the \op:
\begin{equation}
\label{eq:spnorm+_opsmallh}
\Op[{\h[1]}]^c = \of{\frac{12}{L g}\h}^{1/3}.
\end{equation}
The dependence on \h, i.e., the exponent $1/\delta=1/3$, is the same as the one with which at \tcb{} the bulk \op{} responds to a weak bulk field; {\eref{eq:spnorm+_opsmallh}}  agrees  with our numerical data (not shown). 
Accordingly, the free energy (per $k_b T$ and $A$) at criticality is
\begin{equation}
\label{eq:fesmallhcrit}
	\fe[cr]\of{\h[1]=\h[2]\to 0}
	= -\frac{3}{2}\of{\frac{12 \h^{4}}{L g}}^{1/3}.
\end{equation}
Therefore the critical Casimir amplitude $\Delta_+\of{\hs[1]}$ depends on the surface fields as
\begin{equation}
\label{eq:spnorm+_forcesmallh}
\Delta_+\of{\hs[1]} \sim -\hs[1]^{4/3}, \quad  l_{sp}> L.
\end{equation}
This checks with our numerical data shown in  Fig.~\ref{fig:spnorm_casampl-various}(b). Note that in the regime $L<l_{sp}$ the film thickness has not yet reached asymptotically large values. As a consequence $f_{C}\sim L^{-4/3}$ which is a much slower decay than $\sim L^{-d}$ in the asymptotic regime $L\gg l_{sp}$ (\eref{eq:ccfscaling}).

The change from a purely repulsive \fsf{} with a maximum below \tcb{} (which are the characteristics of the \of{+,-} BCs) to a purely attractive \fsf{} with a minimum above \tcb{} (which are the characteristics of the \of{+,+} BCs) occurs at fixed $\hs[1]>0$ for $\hs[2]\to 0^-$, i.e., in the regime where the rescaling scheme is not applicable. We note that in this crossover regime the overall magnitude of \fs{} is much smaller than the one for the \of{+,-} and \of{+,+} BCs (for all curves shown in Fig.~\ref{fig:spnorm_crossover} the rescaling scheme does apply).
As an exemplary case, we show in Fig.~\ref{fig:spnorm_crossover_smallh} $\fs\of{y}$ for $\hs[1]=168$ and various values of \hs[2]. For $\hs[2]=-4.2 \simeq \hs[2,min]\of{y=0}$, so that $\hs[2]/\hs[1]=-0.025$, the maximum of  \fs{} is located at  $y\simeq0$. Upon a further decrease of \abs{\hs[2]} a minimum develops at some positive value $y_{min}$ and \fs{} exhibits two broad maxima of comparable height, one above and one below \tcb. We have found, that the value of the scaling function at the maxima as a function of \hs[2] varies as $\fs\of{y_{max}}\sim\of{\hs[2]}^2$ for $\hs[2]\to 0$. For a further decrease of \abs{\hs[2]} the minimum becomes negative. For all $\hs[2]<0$ the \fsf{} is positive for sufficiently large values \abs{y}. The case of $\hs[2]=0$ can be expressed in terms of the \of{+,+} BCs (see \eref{eq:rescCas1} and Ref.~\cite{note_rescl_sp}).
In Appendix~\ref{sec:app_crossregime} we present two different approaches to describe analytically the variation of $\fs\of{y;\hs[1],\hs[2]}$ in the crossover regime. These approaches deal with films for which the rescaling scheme is not applicably and within suitable ranges of surface fields they capture well the numerical data for the crossover behavior shown in Fig.~\ref{fig:spnorm_crossover_smallh}.

\subsection{Crossover from the ordinary to the normal transition \label{ssc:ordnorm}}

A surface with a nonzero surface enhancement $\c[,1]>0$ and with $\h\neq0$ falls into the crossover regime between the ordinary and the normal transition.
From  the boundary condition in \eref{eq:bc1} it follows that within MFT for a surface field $\h=\c[,1]\Op[b]$ the semi-infinite profile is spatially constant: $\Op\of{z}=\Op[b]$. Here we consider only the case $\h[b]=0^{\pm}$ (see \eref{eq:slitham}) for which $\pm$ coincides with the sign of \h{} (i.e., the phase preferred by the surface is the same as the one prevailing in the bulk) in order to avoid complications induced by wetting transitions which occur if the surface preference is opposite.
Thus for stronger (weaker) surface fields the value of the \op{} at the surface is larger (smaller) than in the bulk. For $\tau>0$, $\Op[b]=0$ and the surface is ordered for any $\h>0$, resembling the normal transition. On the other hand, from \eref{eq:opb} it follows that for temperatures sufficiently below \tcb{}, i.e.,
\begin{equation}
\label{eq:tau_wetting}
	\tau<\tauw[,1]=-\of{g/6}\of{\h[1]/\c[,1]}^2,
\end{equation}
where $\tau_{w,1}$ is given implicitly by $\Op[b]\of{\tau_w}=\h[1]/\c[,1]$ (which happens to be the wetting transition temperature at wall $1$ for $\h[b]=0^{\mp}$ \cite{Parry-et1992}), due to the suppressive influence of \c[,1] the order at the surface is lower than in the bulk and increases with increasing distance from the surface, resembling the \op{} behavior corresponding to the ordinary transition.
Therefore, for films exposed to nonzero and finite surface parameters we can identify the following regimes resembling  different surface UCs: (1) far above \tcb{} the \of{+,\pm} UC (depending on the relative sign of the surface fields)  and  (2)  far below \tcb{} (\eref{eq:tau_wetting}) the \of{ord,ord} UC. For temperatures in between, a regime  resembling  the \of{+,ord} UC can exist. 
The crossover between these different regimes  gives rise to richly structured \fsfs{} (see below).

Since the \ccf{}  is repulsive for films with \of{+,-} and \of{+,ord} BCs and attractive for films with \of{+,+} and \of{ord,ord} BCs, in view of the above discussion the \ccf{} is expected to change its sign up to two times. Let us first consider the case $\h[1]/\c[,1]>\h[2]/\c[,2]>0$ so that $\tauw[,1]<\tauw[,2]$. Accordingly the change from an attractive to a repulsive force is expected to occur at a certain temperature $\tau_{c,2}=\tau_{w,2}+\delta\tau_{2,1}$ upon lowering the temperature (crossover from the regime resembling the \of{+,+} UC to the one resembling the \of{+,ord} UC), followed upon further lowering the temperature by a change from repulsion to attraction again at $\tau_{c,1}=\tau_{w,1}+\delta\tau_{1,2}$ (crossover from the regime of the \of{+,ord} UC to the one of the \of{ord,ord} UC). If $\h[1]/\c[,1]>-\h[2]/\c[,2]>0$ only one change from a repulsive to an attractive force is expected to occur at the temperature $\tau_{c,1}=\tau_{w,1}+\delta\tau_{1,2}$ (crossover from the regime resembling the \of{+,ord} UC to the one resembling the \of{ord,ord} UC), because at $\tau_{c,2}=\tau_{w,2}+\delta\tau_{2,1}$ the film crosses over from the regime of the \of{+,-} UC (due to $\h[1]\h[2]<0$) to the one of \of{+,ord} UC with both UCs rendering the \ccf{} to be repulsive. $\delta\tau_{i,j}$ are the finite-size corrections depending on the properties of both surfaces. The changes of sign and the exact values of $\tau_{c,i}$ will be discussed in more detail in Subsec.~\ref{ssc:signchange}.

In the following we restrict ourselves to a more qualitative description of the features in the crossover regime, although for particular regimes the rescaling and the perturbation theory, as introduced in Subsec.~\ref{sss:rescale} and in Appendix~\ref{sec:app_crossregime}, respectively, should be applicable as well.

First we discuss the symmetric case $\cs[1]=\cs[2]$ and $\hs[1]=\hs[2]$.
In Fig.~\ref{fig:ccf_ordnorm_symm}  we show  our numerical data for the \fsfs{} 
$\fs[symm]\of{y;100,\hs[1]} \equiv \fs\of{y;\cs[1]=\cs[2]=100,\hs[1]=\hs[2]}$.
We note that  \fs[symm] is negative for all values of $y$ as in the case of the \of{+,+} and \of{ord,ord} BCs. It exhibits a minimum above \tcb, like the fixed point \sfct{} \fs[\of{+,+}] \cite{Krech1997}.
This minimum  is very shallow for weak \hs[1] and, as expected, deepens for stronger surface fields \hs[1]. As discused above, for a more negative scaling variable $y$ the film crosses over to the asymptotic regime of the \of{ord,ord} BCs. This results in the appearance of an additional minimum below \tcb, as it occurs for the \sfct{} \fs[\of{ord,ord}] \cite{Maciolek-et:2007}. This minimum deepens for decreasing \hs[1]. For finite and nonzero values of the surface parameters \of{\cs[1],\hs[1]} the cusp-like minimum of \fs[\of{ord,ord}] (which is a MFT artefact \cite{Maciolek-et:2006,Maciolek-et:2007}) is smeared out (see Fig.~\ref{fig:ccf_ordnorm_symm}). For $\cs[1]=100$ the two minima are equally deep for $\hs[1]=168$. Between the two minima \fs[symm] exhibits a maximum at the value $y=y_{w,1}$, corresponding to \tauw[,1] (\eref{eq:tau_wetting}), with $\fs[symm]\of{y_{w,1}}=0$ for all $\hs[1]=\hs[2]$ and $\cs[1]=\cs[2]>0$. (For very weak (very strong) \hs[1] the minimum above (below) \tcb{} is very shallow and finally disappears for $\hs[1]\to0$ ($\hs[1]\to\infty$) and thus also the maximum in between becomes hardly detectable.)

For general values of the surface parameters, the crossover values \yc[,1] and \yc[,2], corresponding to \tauch[,1] and \tauch[,2], respectively, differ and, as discussed above, the film crosses through an additional regime corresponding to the \of{+,ord} BCs. This is illustrated in Fig.~\ref{fig:ccf_ordnorm_hr} for the case $\cs[1]=\cs[2]=100$, $\hs[1]=168$, and different \hs[2].

For large surface fields with opposing sign, $\hs[1]\hs[2]<0$, the system is in the asymptotic regime of the \of{+,-} BCs, i.e., the \ccf{}  is repulsive and exhibits a maximum  below \tcb{} (see the curve for $\hs[2]/\hs[1]=-4$ in Fig.~\ref{fig:ccf_ordnorm_hr}). With decreasing $\abs{\hs[2]}$, according to \eref{eq:tau_wetting} the asymptotic regime of the ordinary transition is reached at the second surface already for less negative values of $y$:  the \fsf{} becomes negative at negative values of $y$ and exhibits a minimum there (see the curves for $\hs[2]/\hs[1]\in\set{-2,-0.9}$ in Fig.~\ref{fig:ccf_ordnorm_hr}). For the case of fixed $\cs[1]=\cs[2]=100$ and $\hs[1]=168$ the minimum is deepest for $\hs[2]\simeq-0.9\hs[1]$. 

For $\hs[2]=0$ the \fsf{} is positive for $y>\yc[,1]\simeq-3.6$ and exhibits a maximum $\fs\of{y_{max}}/\ccapp\simeq0.025$ at $\yc[,1]<y_{max}\simeq-1<0$, like for the \of{+,ord} BCs (see Fig.~\ref{fig:ccf_ordnorm_hr}). For $y<\yc[,1]$ the film is in the asymptotic regime of the \of{ord,ord} BCs and the \fsf{} is negative with a minimum $\fs\of{y_{min}\simeq-14}/\ccapp\simeq-0.114$.

For $\hs[1]\hs[2]>0$ and for large positive and very negative values of $y$ the film is in the asymptotic regime of the \of{+,+} BC and the \of{ord,ord} BC, respectively, and thus \fs{} is negative in these limits. However, for large values of \hs[2], \abs{\yc[,2]} becomes very large such that the limit of negative \fs{} is not seen anymore (see the curve for $\hs[2]=10\hs[1]$ in Fig.~\ref{fig:ccf_ordnorm_hr}). In the intermediate regime between \yc[,1] and \yc[,2], corresponding to the \of{+,ord} BCs, \fs{} is positive with a maximum at negative $y$ (see the curves for $\hs[2]/\hs[1]\in\set{2,10}$ in Fig.~\ref{fig:ccf_ordnorm_hr}).

Close to the ordinary transition there is only one relevant, linear scaling field $\hxs[i]{ord}=\hs[i]/\cs[i]$ associated with a single surface (\eref{eq:ordscalfield}). As expected, we see in our data that the \sfcts{} $\fs\of{y}$ for different values of \cs[i] and \hs[i] but fixed $\hxs[i]{ord}=\hs[i]/\cs[i]$ indeed exhibit similar behaviors. For small \cs[i] and \hs[i], the details of the shape and the amplitude of \fs{} still depend on their particular values, but with increasing values of these two surface parameters a convergence of the corresponding \fs{} is seen.

In order to illustrate this point, in Fig.~\ref{fig:ccf_ordnorm_cr} we discuss the case  in which \cs[2] is varied at fixed $\cs[1]=100$ and $\hs[1]=\hs[2]=168$.
For $\cs[2]>0$ and large positive (negative) $y$ the film corresponds to the \of{+,+} BCs (the \of{ord,ord} BCs) and \fs{} is negative with a minimum above (below) \tcb{} (for $\cs[2]/\cs[1]\gtrsim 50$ partly not visible on the scale of the figure). In between there is an intermediate regime, corresponding to the \of{+,ord} BCs where \fs{} is positive with a maximum below \tcb. With increasing \cs[2], i.e., surface 2 approaching  the ordinary transition, \fs{} becomes positive already at larger values of $y$ and finally the regime of negative \fs{} for positive $y$ becomes hardly visible (see the curve for $\cs[2]=50\cs[1]$). On the other hand, the minimum at negative $y$ deepens with increasing \cs[2]. We point out, that this is the same qualitatively behavior as for fixed \cs[i] and $\hs[2]/\hs[1]\to 0^+$ (compare the discussion above and Fig.~\ref{fig:ccf_ordnorm_hr}). For decreasing \cs[2] the minimum at positive $y$, corresponding to the \of{+,+} BCs regime, depeens (compare the curves for $\cs[2]/\cs[1]\in\set{5,1,0.5,0}$ in Fig.~\ref{fig:ccf_ordnorm_cr}). The regime corresponding to the \of{ord,ord} BCs with negative \fs{} finally disappears for $\cs[2]\to 0$ (see Fig.~\ref{fig:ccf_ordnorm_cr}). The same behavior is seen for increasing \hs[2], since it leads to a stronger ordered surface as decreasing \cs[2] does (compare the curve for $\hs[2]/\hs[1]=10$ in Fig.~\ref{fig:ccf_ordnorm_hr} and the curve for $\cs[2]/\cs[1]=0$ in Fig.~\ref{fig:ccf_ordnorm_cr}).

For a negative surface enhancement $\cs[2]<0$ the trends of deepening of the minimum above \tcb{} and increasing of the maximum below \tcb{} continues (see the curve for $\cs[2]/\cs[1]=-0.1$ in Fig.~\ref{fig:ccf_ordnorm_cr}), because the extraordinary transition, which corresponds to $\cs[2]<0$, and the normal transition are equivalent (see Subsec.~\ref{ssc:surfuni}).

\subsection{Change of sign of the critical Casimir force \label{ssc:signchange}}

In the crossover regime we observe the interesting feature that the \sfct{} \fs{} of the \ccf{} changes sign as function of the scaling variable $y=t\of{L/\xi_0^+}^{1/\nu}$. This implies that at fixed temperature the \ccf{} changes from being attractive to being repulsive (or reverse) upon varying the distance between the two surfaces. Equivalently this change also occurs at a fixed distance upon varying the temperature. Thus temperature allows one to control both the strength and the sign of the \ccf.

In order to understand and to interpret this change of sign we consider the functional form of the \opp{} $\Op[semi]\of{z\geq -L/2;\tau,\Sf[1]}$ in the semi-infinite system with the surface located at $z=-L/2$ and with the surface parameters $\Sf[1]=\of{\c[,1],\h[1]}$; this is available in the literature, e.g., in Ref.~\cite{{Lubensky-et1975}}. Using the BC (for $z=L/2$) in \eref{eq:bc2}, for certain values of the parameters $\Sf[2]=\of{\c[,2],\h[2]}$ of a second surface located at $z=L/2$ one can find a temperature $\tau=\tauch$ such that
\begin{equation}
\label{eq:sf2prop}
\h[2]=\c[,2]\Op[semi]\of{z=L/2;\tauch,\Sf[1]} +\Op[semi]'\of{z=L/2;\tauch,\Sf[1]}.
\end{equation}
Upon construction, at this temperature $\tauch$, $\Op[semi]\of{z;\tauch,\Sf[1]}$ coincides  with the profile in a film (of width $L$ and with confining walls \Sf[1] and \Sf[2]) for distances $-L/2\leq z\leq L/2$. Equation~\eqref{eq:sf2prop} implicitly defines classes $\mathcal{C}\of{L,\tauch; \Sf[1] }$ of walls \Sf[2] which do not disturb the semi-infinite profile if they are inserted at the distance $L$ and at the temperature $\tauch$.
Since such OP profiles of finite and semi-infinite systems coincide, the \ccf{}  for a film of thickness $L$ with surfaces \Sf[1] and $\Sf[2]\in\mathcal{C}\of{L,\tauch; \Sf[1]}$ at this temperature $\tauch$ is zero.
Because in the semi-infinite geometry the \op{} varies algebraically for $\abs{z+L/2}\ll \xi$ and decays exponentially towards its bulk value for $\abs{z+L/2}\gg\xi$ (see, e.g., Ref.~\cite{diehl1997}), for sufficiently large  $L$ the value of the \op{} at the position of the second "nondisturbing" surface $\Sf[2]\in{\mathcal{C}\of{L,\tauch; \Sf[1] }}$ and for the temperature $\tauch$ is close to its bulk value. Thus, in a first approximation, the change of sign occurs if at one of the surfaces the \op{} takes its bulk value, $\Op[i]=\Op[b]$. We recall, that a surface which enhances the order, i.e., at which $\abs{\Op[i]}>\abs{\Op[b]}$, corresponds to the normal transition \of{\pm} while a surface that suppresses the order, i.e., at which $\abs{\Op[i]}<\abs{\Op[b]}$, corresponds to the ordinary transition \of{ord} and that the \ccf{} is attractive (repulsive) for \of{+,+} and \of{ord,ord} BCs (for \of{+,-} and \of{+,ord} BCs).

Based on these features, within MFT the sign of the \sfct{} \fs{} of the \ccf{} can be inferred rather reliably from the values $\Ops[1]\equiv\Ops\of{\zs=-0.5;y,\cs[1],\hs[1],\cs[2],\hs[2]}$ and $\Ops[2]\equiv\Ops\of{\zs=0.5;y,\cs[1],\hs[1],\cs[2],\hs[2]}$ of the scaled \op{} at the surface 1 and 2, respectively:
\begin{equation}
\label{eq:signccf_and_op}
\begin{split}
	\fs\of{y;\cs[1],\hs[1],\cs[2],\hs[2]}<0&\;
	\text{if }
		\begin{cases}
		\Ops[b]\of{y} < \Ops[1],\Ops[2] \text{ or}\\
		\Ops[b]\of{y} > \Ops[1],\Ops[2]
		\end{cases} 
	\\
	\fs\of{y;\cs[1],\hs[1],\cs[2],\hs[2]}>0& \;
	\text{if }
		\begin{cases}
		\Ops[2] < \Ops[b]\of{y} < \Ops[1] \text{ or}\\
		\Ops[2] > \Ops[b]\of{y} > \Ops[1]
		\end{cases} .
\end{split}
\end{equation}

In Fig.~\ref{fig:signchange} this is illustrated for two examples. 
The case $\cs[1]=\cs[2]=0$, $\hs[1]=2.5$, and $\hs[2]=-0.1\hs[1]$ is representative for the crossover regime between the normal and the special transition, like the cases shown in Fig.~\ref{fig:spnorm_crossover_smallh}; note, however, that there $\hs[1]=168$ so that \fs{} differs even if the ratio $\hs[2]/\hs[1]$ is the same.
The \sfct{} $\fs\of{y;\cs[1]=\cs[2]=0,\hs[1]=-10\hs[2]=2.5}$ is positive for $\abs{y}\gtrsim5$ and decays to zero for $\abs{y}\to \infty$ (see Fig.~\ref{fig:signchange}(a)). It exhibits maxima at $y_{max,1}\simeq-4$ and $y_{max,2}\simeq9$ of approximately equal height $\fs\of{y_{max,i}}/\ccapp \simeq 0.00066$. Between these two maxima \fs{} decreases towards negative values  and exhibits  a minimum $\fs\of{y_{min}\simeq1}/\ccapp \simeq -0.0049$. 
As shown in the inset of Fig.~\ref{fig:signchange}(a), for $y\gtrsim9$ the relation  $\Ops[1]>0=\Ops[b]\of{y>0}>\Ops[2]$ holds and the \ccf{}  is repulsive. For $y\simeq8$, \hs[2] is too weak to overturn the positive order imposed by \hs[1] and \Ops[2] turns positive.
For $y<8$ the criterion  proposed in \eref{eq:signccf_and_op} predicts an attractive \ccf,
whereas the actual change of sign occurs at $y\simeq5$. This discrepancy occurs because  the \ccf{}  is zero if $\Op[2]=\Op[semi]\of{z=L/2,\tau,\Sf[1]}$ whereas in \eref{eq:signccf_and_op}, which is proposed to hold for large $L$,  we assume  $\Op[semi]\of{z=L/2,\tau,\Sf[1]}\approx \Op[b]$.
Below \tcb, the bulk value of the order parameter is nonzero (and positive, see \eref{eq:scaledopb} and the horizontal dotted lines in the insets of Fig.~\ref{fig:signchange}). For decreasing, i.e., more negative values of  $y$, the increase of the positive order at the second surface \of{\cs[2],\hs[2]} is reduced by the opposing surface field $\hs[2]<0$, whereas at the first surface \of{\cs[1],\hs[1]} the positive order is enhanced. At approximately $y\simeq -3$ the bulk order \Ops[b] and the surface order \Ops[2] become equal and for $y\lesssim-3$ the relation $0<\Ops[2]<\Ops[b]<\Ops[1]$ for the \op{} holds. Hence, according to \eref{eq:signccf_and_op} a repulsive \ccf{}  is expected. As can be seen in Fig.~\ref{fig:signchange}(a) this is indeed the case, but the change of sign occurs at $y\simeq -1.5$. Again, this discrepancy occurs due to the asymptotic character of our simplified criterion in \eref{eq:signccf_and_op}.

As a second example Fig.~\ref{fig:signchange}(b) shows the \fsf{} for the case $\cs[1]=\cs[2]=100$ and $\hs[1]=10\hs[2]=1680$ corresponding to the crossover between the normal and the ordinary transition.
The \fsf{} is positive for $y\lesssim -1.5$ and negative otherwise. In general, for strong surface parameters \cs[] and \hs[] considered here the values of the \op{} at the surfaces do not change significantly upon varying the scaling variable $y$ (compare the inset of Fig.~\ref{fig:signchange}(b)). For $50>y>-50$ the values of the \op{} at the surfaces are in the range $14.9<\Ops[1]<15.5$ and $1.5<\Ops[2]<2$. As in the first example shown in Fig.~\ref{fig:signchange}(a), the change of sign of the \fsf{} occurs approximately at that (here single) value of $y$ for which $\Ops[b]\of{y}$ equals \Ops[2] (see Fig.~\ref{fig:signchange}(b)). Due to the strong surface field \hs[1], \Ops[1] is larger than \Ops[b] for the whole range of $y$ within which the \fsf{} has a detectable amplitude. 
Far away from criticality the correlations reduce to microscopic length scales and the \ccf{}  becomes vanishingly small. Thus in the case discussed here no attractive \ccf{}  is observed for very negative values of $y$.

\section{Comparison between $d=2$, $d=3$, and $d=4$ \label{sec:dim}}

\subsection{$d=2$ and $d=4$ }
Recently \cite{Abraham-et:2009,Nowakowski-et:2008,Nowakowski-et:2009} the crossover behavior of the \ccf{}  has been studied for a two-dimensional Ising strip  in the limit of $M\gg1$, where $M$ is the number of rows of the strip, i.e., its thickness. These authors considered the case of an unchanged coupling constant $J_{\parallel}$ in the two surface rows of the strip, 
\begin{equation}
\label{eq:dim_coupling}
J_{\parallel}=J\of{1+\Delta_{\parallel}}, \qquad \Delta_{\parallel}=0, 
\end{equation}
where $J$ is the nearest-neighbor coupling constant in the bulk.
Carrying out a systematic continuum limit one can relate the lattice parameters and the couplings in the continuum model.
The surface enhancement \c{} is related to $\Delta_{\parallel}$ \cite{Binder1983,Diehl1986,Lubensky-et1975}:
\begin{equation}
\label{eq:relate_c_and_Delta}
\c = \of{ 1-2\of{d-1}\Delta_\parallel}/a,
\end{equation}
where $a$ is the spacing of the simple cubic lattice. For $\Delta_\parallel<\of{2\of{d-1}}^{-1}$ (within MFT) the continuum limit $a\to 0$ leads to $\c=\infty$ which corresponds to the ordinary transition. The scaling variables $x=sign\of{t}\abs{t}^{\nu}M/\xi_0^+$ and $z\sim\h[1]^{\nu/\Delta_1}M$, with $\nu\of{d=2}=1$ and $\Delta_1^{ord}\of{d=2}=\Delta_1\of{d=2}=1/2$, as introduced in Ref.~\cite{Nowakowski-et:2009} translate into the the scaling variables we use here according to $x=y^{\nu}$ and $z=\of{\hxs[1]{ord}}^{\nu/\Delta_1}$ (compare Eqs.~\eqref{eq:y} and \eqref{eq:opp_scaling_2} and the expression after \eref{eq:ordexp}).
Figure~9(a) in Ref.~\cite{Nowakowski-et:2009} corresponds to the symmetric case and thus corresponds to Fig.~\ref{fig:ccf_ordnorm_symm} here. The qualitative behavior is the same in both cases. The \sfct{} \fs{} (called $\tilde{\mathcal Y}$ in Ref.~\cite{Nowakowski-et:2009}) is negative for the whole range of the scaling variable $y$. For weak surface fields, \fs{} exhibits a minimum below \tcb{} which becomes more shallow upon increasing \hs[1], while a minimum above \tcb{} develops concomitantly and the absolute value of its depth increases. In contrast to $d=4$, where the amplitude of the maximum between the two minima vanishes ($\fs\of{y_{max};d=4}=0$), in $d=2$ the amplitude of the maximum is nonzero (see, e.g., the dash-dotted curve for $z=0.5$ in Fig.~9(a) in Ref.~\cite{Nowakowski-et:2009}).
Concerning the antisymmetric case in $d=2$ (see Fig.~9(b) in Ref.~\cite{Nowakowski-et:2009}) and $d=4$ (available data for $\hs[1]>0$ and $\hs[2]/\hs[1]=-1$ are not shown, but the qualitative behavior can be inferred from the curves for $\hs[2]/\hs[1]\in\set{-0.9,-2}$ in Fig.~\ref{fig:ccf_ordnorm_hr}) \fs{} has one maximum with $\fs[max]>0$ and one minimum with $\fs[min]<0$. In both cases, the minimum shifts towards more negative values of $y$ for increasing \hs[1]. For $d=2$ the position of the maximum is located at positive $y$ for small \hs[1], moves to smaller values of $y$ with increasing \hs[1], and for a particular value of \hs[1] it is located at $y=0$. In contrast, for $d=4$ and within the ranges of values of \cs[1] and \hs[1] for which the value of the maximum is larger than numerical errors, the maximum is always located at negative values of $y$. 
In  Refs.~\cite{Nowakowski-et:2008,Nowakowski-et:2009} the authors used also different scaling variables. We have compared the resulting \sfcts{} expressed in those variables, too, and have also found the same trends in $d=2$ and $d=4$.

In  Ref.~\cite{Nowakowski-et:2008} and in its extended version \cite{Nowakowski-et:2009} only  the cases $\hs[2]=\pm\hs[1]$ were considered.  Results for the \ccf{} in the two-dimensional Ising film subject to {\it arbitrary} surface fields and for $\Delta_{\parallel}=0$ (\eref{eq:dim_coupling}) are provided in Ref.~\cite{Abraham-et:2009}. All three studies for $d=2$ show, that the \ccf{} can change sign not only by changing the surface fields but also by varying the temperature. In the crossover regime the \fsf{} exhibits more than one extremum and thus the behavior in $d=2$ is in qualitative agreement with the present MFT results.

\subsection{$d=3$ and $d=4$ \label{ssc:3d}}
Due to a dearth of analytical means, the natural choice for studying the \ccf{} in $d=3$ consists of Monte Carlo (MC) simulations for a three-dimensional Ising film on a simple cubic lattice with \of{100} surfaces \cite{MC}.  Within this approach, the surface fields are mimicked by additional layers in which all spins are set equal to $1$ or $-1$ and which couple to the actual surface layers of the film by a modified coupling constant $J_{\perp}=\alpha J$, with $J$ as the nearest neighbor coupling constant in the bulk ($\alpha=J_{\perp}/J$ measures the modified coupling constant $J_{\perp}$ in units of $J$, not to be confused with the critical exponent $\alpha$). For details concerning the determination of the \ccf{} via such MC simulations see Refs. \cite{Vasilyev-et:2010,MC}.
The universal critical Casimir amplitude for the $d=3$ Ising film is $\Delta_{\of{+,+}}\of{d=3}\simeq 0.38$ \cite{Vasilyev-et:2009}.
The universal critical exponent of the correlation length for the $d=3$ Ising bulk UC is $\nu\of{d=3}\simeq0.63$ \cite{Pelissetto-et:2002} and its non-universal amplitude for the simple cubic Ising model with nearest neighbor coupling is $\xi_0^+\simeq0.50$ \cite{Ruge-et:1994}. 
First results for a three-dimensional Ising film exposed at one surface to a fixed, finite surface field \hs[1] corresponding to $\alpha_1=1$ and at the other one to a variable surface field \hs[2] corresponding to $0\leq\alpha_2\leq1$ are shown in Fig.~\ref{fig:force_3d}.
These results show a similar behavior as the one discussed for $d=4$ in Subsec.~\ref{ssc:ordnorm}:
For $\hs[2]/\hs[1]=1$, i.e., $\alpha_2=1$ the \sfct{} exhibits only one minimum which is located above \tcb{}. With decreasing $\hs[2]/\hs[1]$ (i.e., smaller $\alpha_2$) this minimum becomes shallower and below \tcb{} a maximum develops. For $\hs[2]=0$ (i.e., $\alpha_2=0$) there is no minimum any more. For $d=4$, the disappearance of the minimum above \tcb{} for $\hs[2]=0$ can be followed in Fig.~\ref{fig:ccf_ordnorm_hr}. In the same figure, for $\hs[2]/\hs[1]\in\set{10,2,1}$, one can see that upon decreasing $\hs[2]/\hs[1]$ the minimum becomes shallower. For $d=4$, the scaling functions \fs{} with $0\leq\hs[2]/\hs[1]\leq2$ shown in Fig.~\ref{fig:ccf_ordnorm_hr} exhibit a minimum also below \tcb, which is not seen in the data for $d=3$. In $d=4$, below \tcb{} no minimum can be detected if the   surface field \hs[1] is very strong (see the curves for $\hs[1]=\hs[2]\in\set{500,1000}$ in Fig.~\ref{fig:ccf_ordnorm_symm}). This suggests, that in $d=3$ the surface field \hs[1] corresponding to $\alpha_1=1$ is so strong that surface $1$ stays in the regime of the normal surface UC (compare the discussion in Subsec.~\ref{ssc:ordnorm}) and thus prevents the occurrence of a minimum also below \tcb{}. Accordingly, a minimum below \tcb{} together with one above \tcb{} is expected to occur for certain values $\alpha_1<1$ also in $d=3$. In order to check this expectation, further MC simulation data for these cases are required.

Also for opposing surface fields the same trends appear in $d=3$ and $d=4$ as evidenced by comparing MC simulation data for a film exposed to surface fields corresponding to $\alpha_1=1\geq-\alpha_2\geq0$ \cite{Vasilyev-et:2010} with the present MFT results.

\section{Summary \label{sec:summary}}

\begin{figure}
\includegraphics{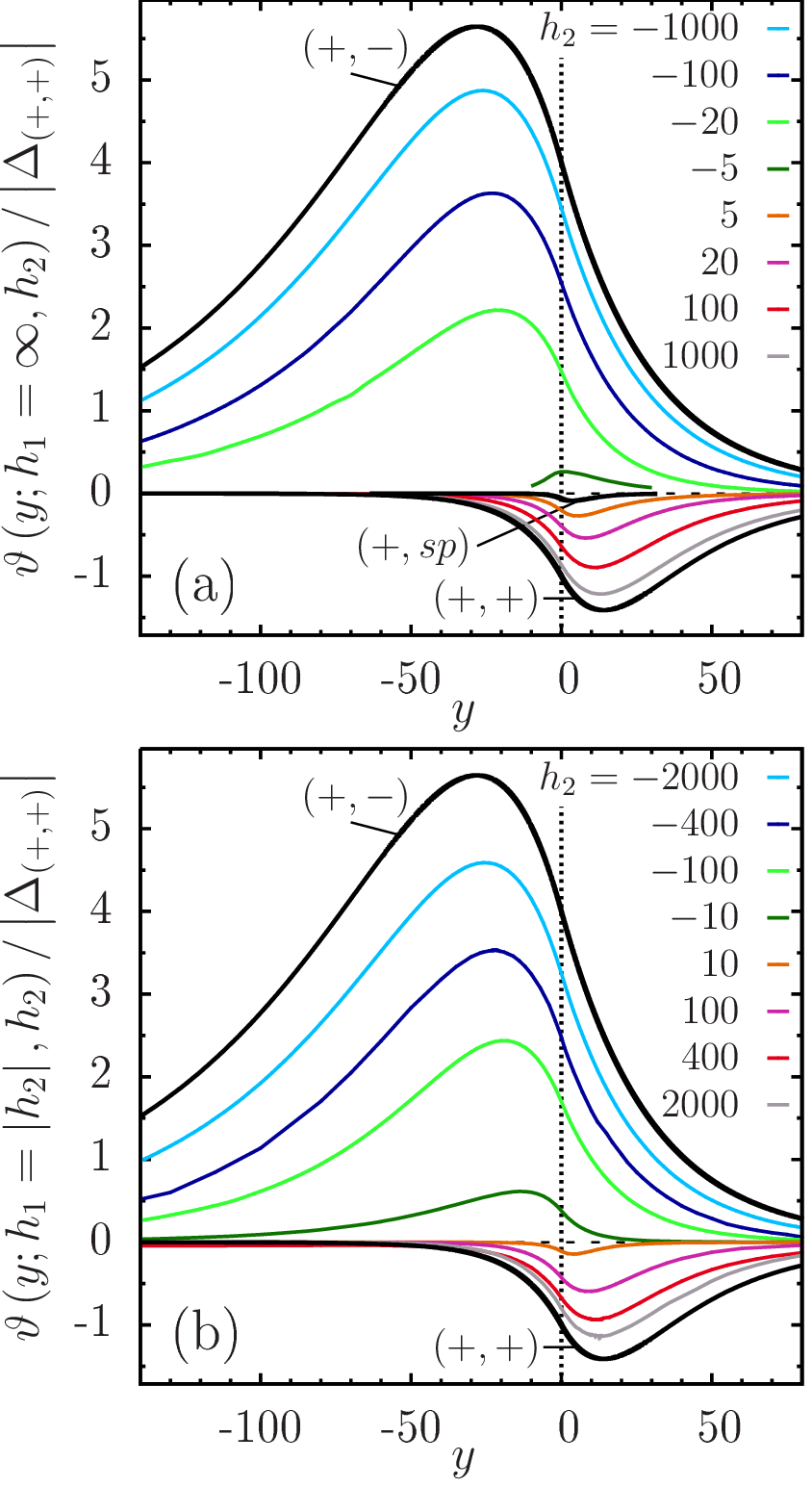}
\caption[spnorm_crossover]
	{The \sfcts{} \fs{} of the \ccf{} are shown for various types of confining surfaces covering the crossover between the normal transition ($\pm$, i.e., $\hs[i]=\pm\infty$, strong adsorption) and the special transition ($sp$, i.e., $\hs[i]=0$, neutral surface, $\cs[1]=\cs[2]=0$). In  (a) $\hs[1]=\infty$ is fixed and $\hs[2]$ varies from $+\infty$ \fd{\of{+,+}} to $-\infty$ \fd{\of{+,-}}. For all cases shown, \fs{} for finite \hs[2] can be expressed in terms of $\fs[\of{+,\pm}]$ by using \eref{eq:rescCas1}. In (b) \fs{} is shown for the symmetric and antisymmetric cases $\hs[1]=\abs{\hs[2]}$ and $-\infty<\hs[2]<\infty$. In general the convergence towards the strong adsorption limit (black lines) is rather slow (compare Fig.~\ref{fig:spnorm_casampl-various}(a)). 
}
\label{fig:spnorm_crossover}
\end{figure}

\begin{figure}
\includegraphics{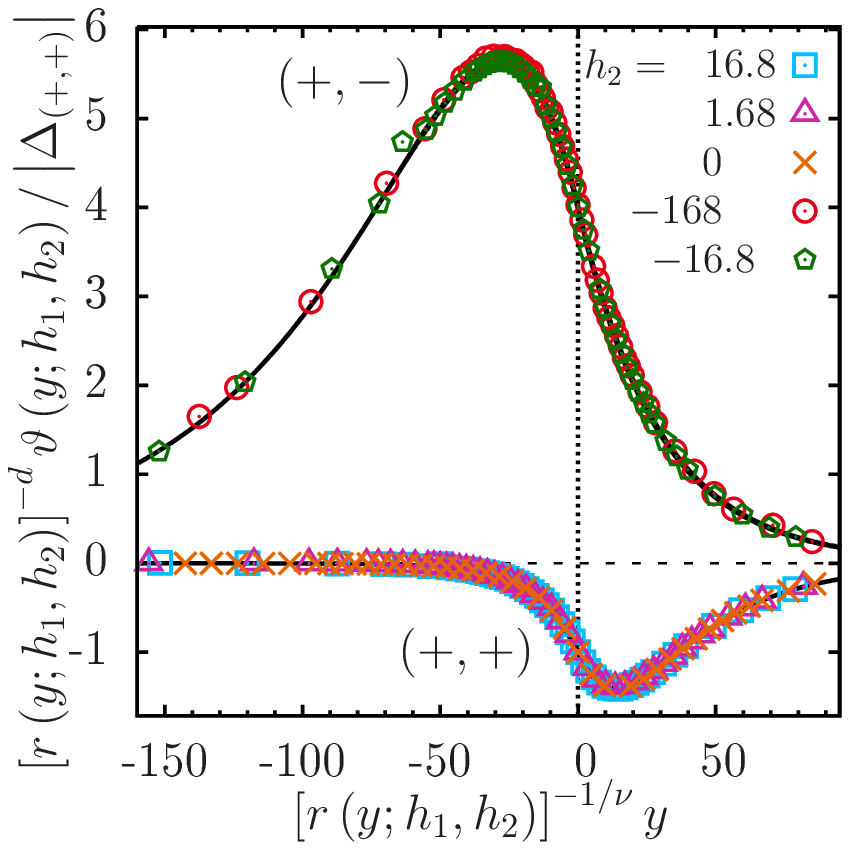}
\caption[rescaling]
	{For all surface fields with $\hs[1]\hs[2]>0$ and certain regimes of surface fields with $\hs[1]\hs[2]<0$ (see \eref{eq:h2min} and the discussion before it) the \sfct{} \fs{} of the \ccf{} for finite surface fields can be mapped onto the fixed point \sfcts{} \fs[\of{+,+}] and \fs[\of{+,-}], respectively: $r^{-4}\fs\of{y;\hs[1],\hs[2]}=\fs[\of{+,\pm}]\of{r^{-2}y}$, $r=r\of{y;\hs[1],\hs[2]}$ (\eref{eq:rescCas1}). The symbols correspond to $\hs[1]=168$ for different values of \hs[2] obtained by minimizing the Hamiltonian in \eref{eq:slitham} and $r\of{y;\hs[1],\hs[2]}$ is given implicitly by Eqs.~\eqref{eq:def_rs1} and \eqref{eq:def_rs2}. \fs[\of{+,+}] and \fs[\of{+,-}] are known analytically \cite{Krech1997} and given by the full black lines. Here $d=4$ and $\nu=1/2$.}
\label{fig:rescaled_force}
\end{figure}

\begin{figure}
\includegraphics{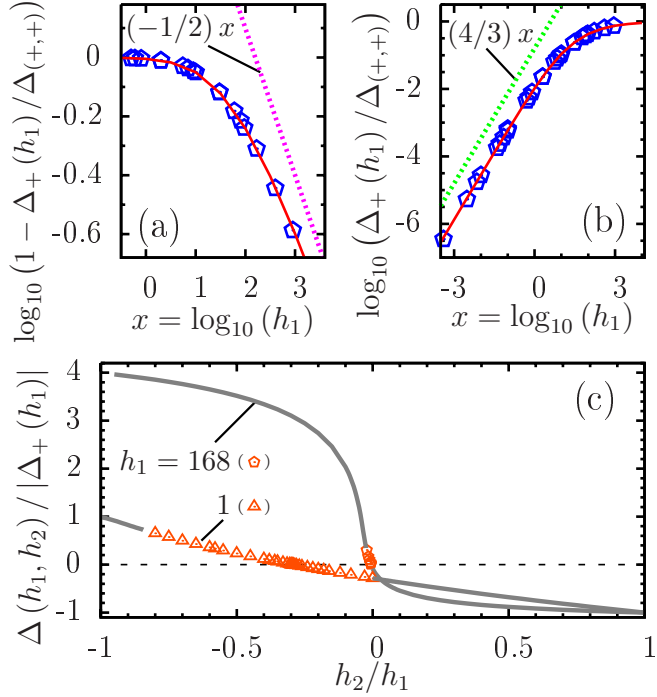}
\caption[Casimir amplitudes]
	{In (a)-(c), symbols represent data obtained from the numerical minimization of the Hamiltonian in \eref{eq:slitham} and the full lines follow from the rescaling scheme (\eref{eq:rescCas1}).
	(a) For $\hs[1]\to\infty$ the critical Casimir amplitude $\Delta_+\of{\hs[1]}=\Delta\of{\hs[1]=\hs[2]}=\fs\of{y=0;\hs[1],\hs[2]}$ approaches its fixed point value $\Delta_{\of{+,+}}$ slowly $\sim\hs[1]^{-1/2}$ from above (see dotted line representing $-(1/2)x+1.1$ and \eref{eq:fsfapproachfix}).
	(b) For small surface fields $\hs[1]=\hs[2]$ and $\cs[1]=\cs[2]=0$ the critical Casimir amplitude $\Delta_+\of{\hs[1]}$ varies $\sim-\hs[1]^{4/3}$ (see the dotted line representing $\of{4/3}x-0.8$ and \eref{eq:spnorm+_forcesmallh}).
	(c) The variation of $\Delta\of{\hs[1],\hs[2]}$ as a function of \hs[2] for fixed $\hs[1]\in\set{1,168}$. The values are normalized by $\abs{\Delta_+\of{\hs[1]}}$ with $\Delta_+\of{\hs[1]=168}/\abs{\Delta_{\of{+,+}}}\simeq-0.51$ and $\Delta_+\of{\hs[1]=1}/\abs{\Delta_{\of{+,+}}}\simeq-0.010$. Figure~\ref{fig:spnorm_casampl-various}(c) focuses on small values of $\hs[2]$ for which the Casimir amplitude changes sign and for which the rescaling property (\eref{eq:rescCas1}) does not hold. In this latter regime the full lines do not apply and are replaced by the numerical obtained data points. These data demonstrate that also in this regime $\Delta\of{\hs[1],\hs[2]}$ varies smoothly. For $\hs[2]/\hs[1]\to 1$ both curves approach $-1$ by construction and due to the fact that for $\hs[1]=\hs[2]$ the Casimir amplitudes are negative. Upon further increasing \hs[2] the Casimir amplitudes turn more negative and approach the limiting values $\Delta\of{\hs[1]=168,\hs[2]=\infty}/\abs{\Delta_{\of{+,+}}}\simeq-0.70$ and $\Delta\of{\hs[1]=1,\hs[2]=\infty}/\abs{\Delta_{\of{+,+}}}\simeq-0.099$. For $\hs[1]=168$ and $\hs[2]/\hs[1]\to-1$ the full line gets close to the MFT value $\Delta_{\of{+,-}}/\abs{\Delta_{\of{+,+}}}=4$ known for $\abs{\hs[i]}=\infty$ \cite{Krech1997}. This property holds also for $\hs[1]=168$ in the limit $\hs[2]\to-\infty$, i.e., $\Delta\of{\hs[1]=168,\hs[2]=-\infty}/\abs{\Delta\of{\hs[1]=168,\hs[2]=\infty}}\simeq4$. For $\hs[1]=1$, $\Delta$ accidently has the same strength for $\hs[2]/\hs[1]=\pm 1$. For more negative values of $\hs[2]\lesssim-\hs[1]$,
}
\label{fig:spnorm_casampl-various}
\end{figure}

We have analyzed the crossover behavior of \ccf[s] for films of thickness $L$ between different surface universality classes. Based on finite-size scaling theory and the numerical solution of the corresponding Landau-Ginzburg theory (\eref{eq:slitham}) we have obtained the following results:
\begin{enumerate}
\item In Fig.~\ref{fig:spnorm_crossover} the \fsf{} $\fs{}\of{y}$ is shown as a function of the scaling variable $y=t\of{L/\xi_0^+}^{1/\nu}$ with $\xi\of{t=\frac{T-\tcb}{\tcb}\to \pm0}=\xi_0^\pm \abs{t}^{-\nu}$ as the bulk correlation length. In Fig.~\ref{fig:spnorm_crossover}(a) $\fs{}\of{y}$ is presented for a film with one strongly adsorbing surface ($\hs[1]=\infty$) and the adsorption preference of the other one continously changing from strongly adsorbing the same component of a binary liquid mixture ($\hs[2]=\infty$) to a 'neutral' surface ($\hs[2]=0$) and to strongly adsorbing the other component ($\hs[2]=-\infty$). The change of the \sfct{} upon varying the strengths of the surface fields for the symmetric ($\hs[1]=\hs[2]$) and the antisymmetric ($\hs[1]=-\hs[2]$) cases is displayed in Fig.~\ref{fig:spnorm_crossover}(b). A slow convergence towards the strong adsorbing limits $\of{+,\pm}\equiv\of{\hs[1]=\infty,\hs[2]=\pm\infty}$ is observed (see also point \ref{item:sum_fsfasymp} below).
\item\label{item:sum_rescaling} The properties of films exposed to finite surface fields $\hs[1]$ and  $\hs[2]$ can be inferred from those of films with $\of{+,\pm}$ fixed point boundary  conditions (BCs) by an effective rescaling (Eqs. \eqref{eq:scalansatz} and \eqref{eq:rescCas1}). This scheme is applicable for all films the surfaces of which prefer the same component ($\hs[1]\hs[2]>0$). For films with surfaces with opposing preferences the scheme is limited to not too asymmetric cases, $\hs[1]\geq-\hs[2]\geq-\hs[2,min]>0$, where $\hs[2,min]$ depends on \hs[1] and the scaling variable $y$. 
In Fig.~\ref{fig:rescaled_force} the successful corresponding rescaling of $\fs{}\of{y;\hs[1],\hs[2]}$ is shown.
\item \label{item:sum_fsfasymp} The proposed rescaling predicts the leading order correction  $\sim\hs^{-1/2}$ for the \ccf{}  relative to the strong adsorption limit $\hs[1]=\infty$ corresponding to \of{+,\pm} BCs (see \eref{eq:fsfapproachfix}). The critical Casimir amplitudes vary accordingly as function of \hs[1] (Fig.~\ref{fig:spnorm_casampl-various}(a)).
\item Symmetric and weakly adsorbing surfaces, $\hs[1]=\hs[2]\to 0$, act like an effective bulk field $\hb=2\hs[1]/L$ (see \eref{eq:fesmallh} and Ref.~\cite{note_dropped_0}). The critical Casimir amplitudes vary $\sim-\hs[1]^{4/3}$ for weak surface fields (\eref{eq:spnorm+_forcesmallh} and Fig.~\ref{fig:spnorm_casampl-various}(b)).
\item The crossover from a purely positive to a purely negative \fsf[,] corresponding to the strong adsorbing limits \of{+,-} and  \of{+,+}, respectively, occurs for the {\it strongly} asymmetric case in the sense that the two surfaces attract different components of the binary liquid mixture, but one does so much weaker than the other, $\hs[1]\gg-\hs[2]>0$. Within this crossover regime \fs{} exhibits two maxima, one above and one below \tcb, and the minimum in between is located above \tcb{} (Fig.~\ref{fig:spnorm_crossover_smallh}). The weaker the second surface field is, i.e., for $\hs[2]\to0^-$, the weaker and broader the maxima are and they are shifted further away from \tcb.
The amplitude of the \ccf{} in this crossover region is one order of magnitude smaller than the maximum value of the \fsf{} for the \of{+,+} BCs and \of{+,-} BCs, respectively.

In Appendix~\ref{sec:app_crossregime} we have put forward two analytic expressions for the \fsf{} in the crossover regime in which the rescaling scheme (see point \ref{item:sum_rescaling}) is not applicable. One (\eref{eq:theta_approx_in_dOp}) is based on a perturbation theory for the \opp{} around the semi-infinite profile. It reproduces nicely the qualitative behavior of \fs{} in the crossover regime and it is in good quantitative agreement for small \hs[2]. 
The other approach shows that the quadratic interpolation of \fs{} as a function of \hs[2] for fixed $y$ and \hs[1], i.e., $\fs\of{\hs[2]}= a+b\hs[2]+c\hs[2]^2$, provides quantitatively reliable results for values $\hs[2,min]<\hs[2]<0$ for which $\fs\of{\hs[2]}$ is not accessible by the rescaling scheme. The coefficients $a$, $b$, and $c$ are obtained by using  $\fs=\fs\of{\hs[2]}$  evaluated at three different values ($0$, \hs[2,s], and $\tilde{\hs[]}_{2}$) of \hs[2] such that $\fs\of{\hs[2]=0}>\fs\of{\hs[2]=\hs[2,s]}=0>\fs(\hs[2]=\tilde{\hs[]}_{2})$ where $\tilde{\hs[]}_{2}\leq\hs[2,min]$.
Figure \ref{fig:spnorm_casampl-various}(c) supports such an interpolation as it shows that the critical Casimir amplitude $\Delta\of{\hs[1],\hs[2]}$ as function of $\hs[2]$ is varying smoothly throughout the whole range of values $\hs[2]/\hs[1]$.
The performance of both approaches is shown in Fig.~\ref{fig:spnorm_crossover_smallh}.
\item For surfaces at which on the one hand ordering is suppressed, e.g., due to missing neighboring liquid molecules at the surface, but where on the other hand there is an adsorption preference, the interplay of these two opposing influences leads to a richly structured $\fs\of{y}$ (see Figs.~\ref{fig:ccf_ordnorm_hr} and \ref{fig:ccf_ordnorm_cr}). This structure can be understood by assuming that for different values of $y$ the system is characterized effectively by different UCs.
This suggests that at different temperatures different properties of the surface exert the dominant influence on the binary liquid mixture.
This is nicely seen for films confined by two identical walls which suppress the order ($\cs>0$) and exhibit an adsorption preference ($\hs[1]>0$). If these two influences are comparable, $\fs\of{y}$ exhibits two minima (see Fig.~\ref{fig:ccf_ordnorm_symm}). The one above \tcb{} corresponds to the adsorption preferences and the one below \tcb{} corresponds to the reduction of the order, as suggested by the comparison with the two limiting cases of pure preference and order reduction, respectively.
This can lead to a situation that the \ccf{} is attractive above \tcb{} and turns repulsive upon lowering the temperature above \tcb{} (see the curve for $\cs[2]/\cs[1]=5$ in Fig.~\ref{fig:ccf_ordnorm_cr}).
\item We have proposed, discussed, and checked a relation between the sign of the \ccf{}  and the values of the \op{} at the surfaces relative to its bulk value (see \eref{eq:signccf_and_op} and Fig.~\ref{fig:signchange}).
For the asymmetric case ($\hs[1]\hs[2]<0$) the sign of the \ccf{} changes above \tcb{} if surface $2$ is not covered anymore by a layer which is enriched in the preferred component. 
In the case that both surfaces prefer the same component ($\hs[1]>\hs[2]>0$) and the effect of missing neighbors is not negligible ($\cs[1],\cs[2]>0$) the \ccf{} changes its sign twice below \tcb, approximately at the two wetting transition temperatures $\tauw[,1]$ and $\tauw[,2]$ (\eref{eq:tau_wetting}) of the two surfaces. This means that the \ccf{} becomes repulsive, if at one surface the attractive influence of the adsorption preference still dominates and at the second one ordering is suppressed relative to the bulk. It switches back to attractive, if at both surfaces ordering is suppressed. If one surface is only very weakly adsorbing, one change of sign is shifted to a temperature above \tcb.
In the asymmetric case ($\hs[1]\hs[2]<0$) the sign changes approximately at the lower wetting temperature, i.e., if at both surfaces the ordering is suppressed (see the inset of Fig.~\ref{fig:ccf_ordnorm_hr}).
Within the present MFT a general, exact, and implicit equation for the temperature at which the change of sign occurs is given in \eref{eq:sf2prop}. Accordingly, the change of sign occurs at that temperature, at which the semi-infinite profile confined by one surface is not disturbed by inserting the other surface. The \ccf{} is attractive, if due to the second surface the ordering is enhanced relative to the semi-infinite profile, i.e., if the film enhances the attraction of the component preferred by the semi-infinite system.
\item Similar behaviors of the \ccf{}  as discussed for the spatial dimension $d=4$ are found in exact results for $d=2$ \cite{Abraham-et:2009,Nowakowski-et:2008,Nowakowski-et:2009} and in Monte Carlo simulation data for $d=3$ \cite{Vasilyev-et:2010} (for $d=3$ see Fig.~\ref{fig:force_3d}). In all three dimensions, for suitably chosen surface properties, the \sfct{} \fs{} of the \ccf{} changes sign as a function of $y$. \fs{} exhibits more than one extrema in the crossover regime and the trends of the position of these extrema are the same in all three spatial dimensions. 
This tells that it should be possible to prepare systems in which the sign of the \ccf{} changes forth and back either by changing the temperature or by changing the film thickness. It is reasonable to expect that this feature translates to systems consisting of colloids the surfaces of which are suitably prepared and which are immersed in critical solvents. Therefore the \ccf[s] provide, at least in principle, a mechanism to prepare stable colloidal suspensions. In Ref.~\cite{Rodriguez} it has also been proposed to stabilize colloidal clusters, however by making use of the omnipresent quantum Casimir forces. The results we presented here suggest that the use of \ccf[s] provides the potential to control via minute temperature changes the distance at which the colloids are in a stable configuration.
\end{enumerate}

\section*{Acknowledgment \label{sec:ack}}
We thank
O. Vasilyev for providing us with not yet published MC simulation data for three-dimensional Ising films.

\appendix

\section{Short distance approximation for the rescaling functions \label{sec:app_rscal-sd}}

Close to the confining walls the variation of the \sfcts{} $\Ops[\of{+,\pm}]\of{\zs}$ of the \opp{} in a film with fixed point \of{+,\pm} BCs can be well approximated by the corresponding \sfct{} for a single wall in the semi-infinite geometry \cite{CA-83}, i.e.,
$\abs{\Ops[\of{+,\pm}]\of{\zs\to\zs[w]}}= \sqrt{2}/\of{\pm\of{\zs-\zs[w]}}$, where  $\zs[w]\in\set{\pm0.5}$ denotes the position of the surface, $\pm\of{\zs-\zs[w]}$ corresponds to $\zs\gtrsim\zs[w]=-0.5$ or $\zs\lesssim\zs[w]=0.5$, and the sign of \Ops{} equals the one of the corresponding (infinite) surface field. If a surface field \hs[i], $i\in\set{1,2}$, is strong enough this relation can be used for that surface $i$ in Eqs.~\eqref{eq:def_rs1} and \eqref{eq:def_rs2} which determine the rescaling functions $r$ and $\zs[0]$. If both surface fields are strong, one obtains
\begin{align}
	r_{sd}^2\frac{\sqrt{2}}{\of{r_{sd}\of{-0.5-\zs[0]^{sd}}+0.5}^2}&=\abs{\hs[1]}&
	\label{eq:app_sd-rs1}\\
	r_{sd}^{2}\frac{\sqrt{2}}{\of{-r_{sd} \of{0.5-\zs[0]^{sd}}+0.5}^{2}}&=\abs{\hs[2]}.&
	\label{eq:app_sd-rs2}
\end{align}
Equations~\eqref{eq:app_sd-rs1} and \eqref{eq:app_sd-rs2} can be written as $2^{1/4}/\abs{\hs[i]}^{1/2}=-0.5\mp\zs[0]^{sd}+0.5/r_{sd}$, with $\mp\to-$ for $i=1$ and $\mp\to+$ for $i=2$, from which one obtains Eqs.~\eqref{eq:r_sd} and \eqref{eq:zs0_sd}. Within this approximation the contributions of the two surfaces to the rescaling parameters are independent of each other. Moreover, within this approximation Eqs.~\eqref{eq:r_sd} and \eqref{eq:zs0_sd} offer a transparent interpretation of these functions: $\zs[0]$ is a measure of the asymmetry of the surface fields the film is exposed to, and $r$ provides a measure of the deviation from the fixpoint BCs \of{+,\pm}.

\section{Analytic expressions for $\fs\of{y}$ in the crossover regime \label{sec:app_crossregime}}
We propose two analytic expressions for the \fsf{} in the crossover regime, i.e., in which the rescaling scheme is not applicable. One is based on a perturbation theory of the \opp{} around the semi-infinite profile. The other approach is the quadratic interpolation of \fs{} as a function of \hs[2] for fixed $y$ and \hs[1].

\subsection{Perturbation theory for the critical Casimir force \label{ssc:app_perturbation}}
In the crossover regime the \sfct{} \fs{} of the \ccf{} exhibits a rich structure with the emergence of three extrema and up to two changes of sign. In order to capture these features qualitatively and quantitatively by an {\it analytic} expression, we approximate the actual \opp{} \Op{} by a term \Op[0] which satisfies the ELE (\eref{eq:ele}), but in general {\it not} the BCs in Eqs.~\eqref{eq:bc1} and \eqref{eq:bc2}, and a perturbation part \dOp{}:
\begin{equation}
\label{eq:opp_ansatz_spnormcross}
\Op\of{z}=\Op[0]\of{z}+\dOp\of{z}.
\end{equation}
Inserting the ansatz of \eref{eq:opp_ansatz_spnormcross} into \eref{eq:ele} provides the differential equation determining the deviation $\dOp=\Op-\Op[0]$, i.e.,
%
%
\begin{equation}
\label{eq:ode_deltaOp}
\dOp''=\of{\tau+\frac{g}{2}\Op[0]^2}\dOp+\hot{{\dOp}^{2}},
\end{equation}
where $\dOp'=\partial\dOp/\partial z$ and \hot{{\dOp}^{2}} stands for terms quadratic and cubic in \dOp{}.
The BCs are (compare Eqs.~\eqref{eq:bc1} and \eqref{eq:bc2}, for simplicity we consider here and throughout this appendix only the crossover from the special to the normal transition, i.e., $\cs[i]=0$)
\begin{align}
        \at{\diff{\dOp}{z}}{z=-L/2} &= -\h[1] - \Op[0]'\of{-L/2} \equiv -\Delta\h[1]
                        \label{eq:delta_bc1}
        \intertext{and}
        \at{\diff{\dOp}{z}}{z=L/2} &= \h[2] - \Op[0]'\of{L/2} \equiv \Delta\h[2].
                        \label{eq:delta_bc2}
\end{align}
(If \Op[0] happens to be the full solution, $\dOp=0$ and thus $\Delta\h[i]=0$.)
The \ccf{} is then obtained from the ansatz $\Op[0]+\dOp$ (with $\dOp\of{\tau,\h[i],\fd{\Op[0]}}$ determined by Eqs.~\eqref{eq:ode_deltaOp}-\eqref{eq:delta_bc2}) via the stress tensor (Eqs.~\eqref{eq:stresstensor} and \eqref{eq:ccf_via_stress}):
\begin{align}
\f&=\f\of{\tau,\fd{\Op[0]}}=
         \T \fd{\Op[0]+\dOp} -\T \fd{\Op[b]}
                \nonumber \\
        &=\T \fd{\Op[0]} + \Op[0]'\dOp' -\of{\tau \Op[0]+\frac{g}{6}\Op[0]^3}\dOp
                -\T \fd{\Op[b]}+\hot{\dOp^{2}},
        \label{eq:ccf_in_dOp}
\end{align}
i.e.,  $\f$ is a function of $\tau$ {\it and} a functional of \Op[0]. Here and in the following we do not indicate explicitly the dependence on the surface fields \h[i].
Concerning the \ccf{} {\it off} the critical temperature, in the following we shall neglect the explicit dependence of \f{} on $\tau$ and  assume that it  enters only via the profile \Op[0], i.e., 
$\f\of{\tau\neq0,\fd{\Op[0]\of{\tau}}}\approx \f\of{\tau=0,\fd{\Op[0]\of{\tau}}}$. 

At the temperature \tauch{}, at which the \ccf{} is zero and changes sign, the actual profile \Op{} in the film \of{-L/2\leq z \leq L/2} coincides, in the case $\h[1]>\abs{\h[2]}$ assumed here and in the following, with the corresponding semi-infinite profile $\Op[s]^*\of{z\geq-L/2,\tauch}$ because at this temperature there is no finite-size contribution to the free energy (see Subsec.~\ref{ssc:signchange}).
We are interested in the change of sign and in the crossover regime around \tauch{} where the \ccf{} is small.
For this case and by invoking the above approximation scheme it is reasonable to dispose of the not yet specified \Op[0] such that $\Op[0]\of{\tau}=\Op[s]\of{\tau}$, where for the time being $\Op[s]\of{\tau}\equiv\Op[s]\of{z,\tau;z_0}$ is any (analytically known \cite{Lubensky-et1975}) semi-infinite solution of the ELE; $z_0=z_0\of{\h[1],z_w}$ is a lengthscale which encodes the dependences on the surface field \h[1] and the position $z_w$ of the confining wall. The specification of this $z_0$ will be discussed below.

In order to proceed we now determine \dOp{} explicitly by assuming that it is sufficiently small and, accordingly, we neglect in the following terms \hot{{\dOp}^2}. For reasons of simplicity we restrict ourselves to solving the linearized ELE (\eref{eq:ode_deltaOp}) at bulk criticality.
According to the above approximation scheme one has
$\Op[0]\of{z;\tau=0}=\Op[s]\of{z,\tau=0;z_0}=\sqrt{12/g}\of{z+z_0}^{-1}$, so that the linearized \eref{eq:ode_deltaOp} turns into $\dOp''\of{z}=6(z+z_0)^{-2} \dOp$ with the solution
\begin{equation}
\label{eq:deltaOp_atTc}
\dOp\of{z;\tau=0}= B_1 \of{z+z_0}^{3} + B_2 \of{z+z_0}^{-2}.
\end{equation}
The coefficients $B_1$ and $B_2$ are determined by the BCs in Eqs.~\eqref{eq:delta_bc1} and \eqref{eq:delta_bc2}:
\begin{align}
        B_1 &= \frac{1}{3} \of{\zp^{-3}\Delta\h[1]+\zm^{-3}\Delta\h[2]}B
                \label{eq:B1}
        \intertext{and}
        B_2 &= \frac{1}{2} \of{\zp^{2}\Delta\h[1]+\zm^{2}\Delta\h[2]}B,
                        \label{eq:B2}
\end{align}
with $B=\zp^{3}\zm^{3}\of{\zp^5-\zm^5}^{-1}$ and $z_{\pm}=z_0\pm L/2$.

The value of the stress tensor for $\Op[0]=\Op[s]$ equals its bulk value, i.e., $\T\fd{\Op[s]} -\T \fd{\Op[b]}=0$, and within the linearized theory (i.e., neglecting terms \hot{\dOp^2}) the \ccf{} at $\tau=0$ is approximated by (see \eref{eq:ccf_in_dOp})
\begin{equation}
\label{eq:cca_in_dOp}
	\f\of{\tau=0,\fd{\Op[s]}}=-5\sqrt{12/g} B_1.
\end{equation}
Straightforward calculations show that \f{} does not depend on $B_2$.

We now resume the remaining task to specify the value of the parameter $z_0$.
We have chosen the straightforward and simplest condition $\Delta\h[1]=0$, i.e., we consider for $\Op[0]=\Op[s]$ the semi-infinite profile corresponding to surface $1$. The condition $\Delta\h[2]=0$ is not considered, because we have observed that the former choice yields better results and it ensures that \Op[s] is non-singular throughout the film for all finite values of $\hs[i]$, $i=1,2$.
Within the crossover regime the profiles $\Op[s]+\dOp$ obtained in such a way compare well with the ones obtained by the full, numerical minimization of \Ham{}, while the critical Casimir amplitude $\Delta$ as function of \hs[2] can be approximated in this way only for values $\hs[2]\approx\hs[2,s]$ (where $\Delta=0$ for $\hs[2]=\hs[2,s]$).
(The full MFT profile \Op{} minimizes the Hamiltonian \Ham{}. Therefore  a more
sophisticated way to determine $z_0$ would be to treat it as a variational parameter
and taking that value $z_0=z_{0,m}$ which minimizes $\Ham\of{\Op[s]\of{z;z_0}+\dOp\of{z;z_0}}$.
)

In order to extend \eref{eq:cca_in_dOp} to values $\tau\neq0$, we apply the approximation stated
below \eref{eq:ccf_in_dOp}, i.e., we take into account only that dependence on $\tau$  which enters via the profile \Op[0]. To this end we express the right hand side of \eref{eq:cca_in_dOp} in terms of the profile $\Op[s]\of{z;\tau=0}$. This can be done by  using the relations $\Op[s,\of{1,2}]\of{\tau=0}= \sqrt{12/g} z_{\mp}^{-1}$ and $\Op[s,\of{1,2}]'\of{\tau=0}=-\sqrt{12/g} z_{\mp}^{-2}$, where $\Op[s,i]=\Op[s]\of{z=z_0\mp L/2;\tau=0}$ is the value of the profile $\Op[0]=\Op[s]$ at the surface $i$ and $\mp\to-$ for $i=1$ and $\mp\to+$ for $i=2$.
Expressing $B_1$  in terms of $\Op[s,i]$ and $\Op[s,i]'$ is not unique, yet it is restricted by the condition to preserve the relation $f_C\sim g^{-1}$ (see \eref{eq:hamscaled} and the note after \eref{eq:ccfscaling}) and to preserve the symmetry with respect to interchanging $1$ and $2$.  

We have found empirically that the final result is {\it not} sensitive to the particular way
of expressing  $B_1$  in terms of $\Op[s,i]$ and $\Op[s,i]'$  and 
in the following we choose a replacement in which  $\Op[s,i]'$ replaces only the terms $z_\pm^2$ in the denominator of $B=1/\of{\zp^2\zm^{-3}-\zm^2\zp^{-3}}$.
Together with Eqs.~\eqref{eq:delta_bc1} and \eqref{eq:delta_bc2} and in terms of the scaling variables given by Eqs.~\eqref{eq:zeta}-\eqref{eq:hs} and \eqref{eq:ccfscaling} this choice leads to the approximation
\begin{equation}
\label{eq:theta_approx_in_dOp}
\fs\of{y;\hs[1],\hs[2]}
        \simeq\frac{6}{g}\frac{5}{9}
                \frac{\of{\hs[1]+\Ops[s,1]^{\prime}}\Ops[s,2]^{3}
                                +\of{\hs[2]-\Ops[s,2]^{\prime}}\Ops[s,1]^{3}} {\Ops[s,2]^{3}\of{\Ops[s,1]^{\prime}}^{-1}-\Ops[s,1]^{3}\of{\Ops[s,2]^{\prime}}^{-1}}.
\end{equation}
$\Ops[s,1]=\Ops[s]\of{\zs=-0.5;y}$ and $\Ops[s,2]=\Ops[s]\of{\zs=0.5;y}$ 
[$\Ops[s,1]^{\prime}=\frac{\partial}{\partial \zs}\Ops[s]\of{\zs=-0.5;y}$ and $\Ops[s,2]^{\prime}=\frac{\partial}{\partial \zs}\Ops[s]\of{\zs=0.5;y}$] are
the values at the two confining surfaces of [the derivative of] the \sfct{} corresponding to a semi-infinite \opp{}:
\begin{equation}
\label{eq:semiinf_prof}
\Ops[s]\of{\zs,y;\hat{\zs}}=\left\lbrace
        \begin{aligned}
        &\of{2y}^{1/2}\fd{\sinh\of{y^{1/2}\of{\zs+\hat{\zs}}}}^{-1}, &
         y>0 \\
        &2^{1/2}\of{\zs+\hat{\zs}}^{-1}, &
         y=0 \\
        &\of{-y}^{1/2}\coth\of{\of{-y/2}^{1/2}\of{\zs+\hat{\zs}}}, &
         y<0
        \end{aligned}
        \right.
\end{equation}
with the parameter $\hat{\zs}$ corresponding to $z_0$ (see above). We have used for $\hat{\zs}$ that value $\hat{\zs}\of{y,\hs[1]}$ which follows from the condition $\Ops[s,1]'=-\hs[1]$ (i.e.,
the one corresponding to that value of $z_0$ as chosen above).
From Fig.~\ref{fig:spnorm_crossover_smallh} we infer that this approximation (\eref{eq:theta_approx_in_dOp}) captures nicely the qualitatively behavior of $\fs\of{y}$ in the crossover regime where three extrema emerge and \fs{} changes sign. 

Within this regime, for small \abs{\hs[2]} it is also in good {\it quantitative} agreement with $\fs\of{y;\hs[1],\hs[2]}$ as obtained from the full numerical minimization of \Ham{} (see the curves for $\hs[2]/\hs[1]\in\set{0,-0.003,-0.007,-0.01}$), whereas for stronger \hs[2] it
underestimates \fs{}. From data not shown we infer that the range of values of
$\hs[2]/\hs[1]$, for which the expression in \eref{eq:theta_approx_in_dOp}
(with $\hat{\zs}$ chosen as above) approximates well $\fs\of{y;\hs[1],\hs[2]}$, increases
with decreasing \hs[1], e.g., for $\hs[1]=0.25$ it captures \fs{} even for $\hs[2]=\pm\hs[1]$. 
A more sophisticated choice for $\hat{\zs}$, which takes into account not only the value
of \hs[1] but also the one of \hs[2], may even improve the value for \fs{} given by \eref{eq:theta_approx_in_dOp} and may extend the range of applicability of this approximation.

\subsection{Interpolation of the scaling function in the crossover regime}

As discussed in the main text, apart from the crossover regime
$0>\hs[2]>\hs[2,min]$ (see \eref{eq:h2min} and the discussion before it, considering here and in the following $\hs[1]>\abs{\hs[2]}\geq0$), the \sfct{}
$\fs\of{y;\hs[1],\hs[2]}$  has the same functional form as \fs[\of{+,\pm}]
(see \eref{eq:rescCas1}) and its value is known analytically, yet implicitly,
because both \fs[\of{+,\pm}] \cite{Krech1997} and the rescaling function
$r=r\of{y;\hs[1],\hs[2]}$ (Eqs.~\eqref{eq:def_rs1} and \eqref{eq:def_rs2}) are
known.
In Appendix~\ref{ssc:app_perturbation} we have provided a perturbation theory for the \sfct{} \fs{} of the \ccf{}. It results in an explicit expression for \fs{} but it is, depending on the value of \hs[1], limited to a certain range of values of \hs[2] and does not necessarily apply
for the whole crossover regime. 

In the following we present a calculation scheme for \fs{} which covers the whole crossover regime and which avoids the full numerical minimization of \Ham{} (\eref{eq:slitham}). 
It turns out that a quadratic interpolation of \fs{} as function of \hs[2] for fixed \hs[1] and $y$, i.e.,
\begin{equation}
\label{eq:fs_crossover}
    \fs^{cross}\of{y;\hs[1],\hs[2]}=a\hs[2]^2+b\hs[2]+c
\end{equation}
adequately serves this purpose. The coefficients $a$, $b$, and $c$ which depend on $y$ and \hs[1] are determined by the condition that $\fs^{cross}$ takes the (exact) values of \fs{} at three distinct values of \hs[2], for which \fs{} is known in terms of analytic expressions, and which we shall specify in the next step. In order to interpolate the value of \fs{} in the crossover regime we take one value in each region where the rescaling scheme applies. As the first value we choose $\hs[2,+]=0$, because it is at the boundary of the crossover regime and it simplifies the calculation of \fs{} (as given by \eref{eq:rescCas1})
because for $\hs[2]=0$ \eref{eq:def_rs2} immediately yields $\zs[0]=0.5$ and one is left with only one implicit equation, i.e., \eref{eq:def_rs1}. As the second value we choose $\hs[2,-]=\hs[2,min]\of{y=0}$ for all values of $y$. This choice avoids that one has to calculate the corresponding $\hs[2,min]\of{y}$ for each value of $y$. (For $\hs[2,-]=\hs[2,min]\of{y\neq0}$ the ensuing interpolation intervall would be narrower than for other choices of \hs[2,-] and thus one would expect in that case more accurate results for $\fs^{cross}$. However it turns out that $\hs[2,min]\of{y=0}$ is sufficiently close to $\hs[2,min]\of{y\neq0}$ to yield satisfactory results; we recall that $\hs[2,min]\of{y=0}\leq\hs[2,min]\of{y}$.)
As the third value we take $\hs[2,s]=\Ops[s]'\of{\zs=0.5;y,\hs[1]}<0$ for which the \ccf{} is zero and thus $\fs\of{\hs[2,s]}=0$ (see \eref{eq:sf2prop}). Since the rescaling procedure amounts to stretching the fixed point \sfct{} \fs[\of{+,\pm}], it cannot describe the qualitatively different shapes of \fs{} in the crossover regime, such as the change of sign of \fs{}. Accordingly, \hs[2,s] lies {\it within} the crossover regime. In sum, the conditions fixing $a$, $b$, and $c$ are:
\begin{subequations}
\label{eq:fs_cross_conditions}
\begin{align}
        \fs^{cross}\of{y;\hs[1],\hs[2]=\hs[2,+]}&=
                \fd{r\of{y;\hs[1],\hs[2,+]}}^{4}\fs[+,+]\of{\fd{r\of{y;\hs[1],\hs[2,+]}}^{-2}y},
        \label{eq:fs_cross_conditions_h2-above}
        \\
        \fs^{cross}\of{y;\hs[1],\hs[2]=\hs[2,-]}&=
                \fd{r\of{y;\hs[1],\hs[2,-]}}^{4}\fs[+,-]\of{\fd{r\of{y;\hs[1],\hs[2,-]}}^{-2}y},
        \label{eq:fs_cross_conditions_h2-below}
        \\
        \fs^{cross}\of{y;\hs[1],\hs[2]=\hs[2,s]}&=0,
        \label{eq:fs_cross_conditions_fs_zero}
\end{align}
\end{subequations}
with \hs[2,\pm] and \hs[2,s] as given above and $r=r\of{y;\hs[1],\hs[2]}$ follows from Eqs.~\eqref{eq:def_rs1} and \eqref{eq:def_rs2}.
The performance of $\fs^{cross}$ (Eqs.~\eqref{eq:fs_crossover} and \eqref{eq:fs_cross_conditions}) is in good agreement with the data obtained by the full, numerical minimization of \Ham{} (\eref{eq:slitham}) (see
Fig.~\ref{fig:spnorm_crossover_smallh}). It is worth mentioning, that a linear interpolation (i.e., $a=0$), using only two of the conditions in \eref{eq:fs_cross_conditions}, performs poorly.
%
%
{}
%


%

\begin{figure}
\begin{quotation}
\noindent
$\Delta\of{\hs[1]=1,\hs[2]}$ first increases. However, upon further decreasing \hs[2], it will decrease and change sign at $\hs[2]=-39.5$ (as obtained from \eref{eq:sf2prop}) and remains negative with a slightly negative limiting value for $\hs[2]\to-\infty$. This can be infered from the fact that $\Delta\of{\hs[1],\hs[2]=-\infty}$ as function of \hs[1] changes sign at $\hs[1,s]=\sqrt{2}$ (Eqs.~\eqref{eq:sf2prop} and \eqref{eq:semiinf_prof}) and is negative for $\hs[1]<\hs[1,s]$.
\end{quotation}
\end{figure}
\cleardoublepage

\begin{figure}
\includegraphics{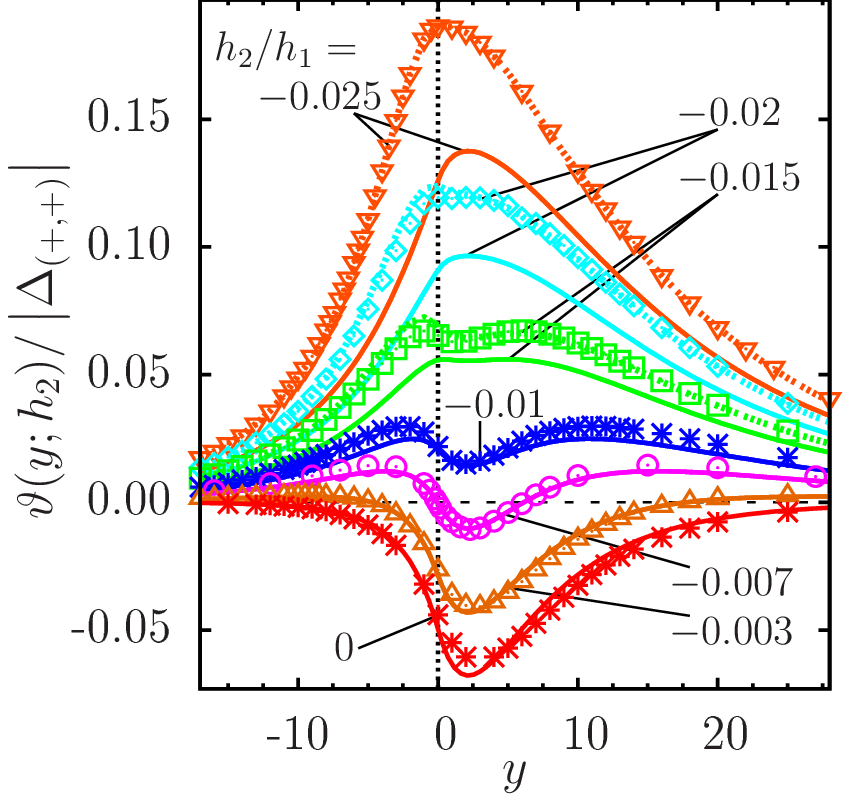}
\caption[small h, crossover regime]
	{The \sfct{} \fs{} of the \ccf{} changes from being purely positive to being purely negative by the formation of three extrema in the crossover regime for $\cs[1]=\cs[2]=0$, fixed $\hs[1]>0$ ($\hs[1]=168$ in the figure), and $\hs[2]\to0^-$. Symbols correspond to data obtained by a full, numerical minimization of the Hamiltonian in \eref{eq:slitham}. For $\hs[1]=168$ the values $\hs[2]/\hs[1]=0,-0.025$ represent the limits of the range of values of \hs[2] for which the rescaling scheme (\eref{eq:rescCas1}) holds (i.e., $\hs[2]/\hs[1]\notin \of{-0.025,0}$). The qualitative behavior of \fs{} for $\hs[2]/\hs[1]\in \fd{-0.025,0}$ is captured by the description obtained from the ELE linearized in terms of the deviation $\dOp=\Op-\Op[s]$ of the \opp{} \Op{} from the profile \Op[s] in a semi-infinite system (see \eref{eq:theta_approx_in_dOp}, full lines). For small \hs[2] it is even in good quantitative agreement (see curves for $\hs[2]/\hs[1]\in\set{0,-0.003,-0.007,-0.01}$). Values obtained by the quadratic interpolation scheme given in \eref{eq:fs_crossover} provide a very good description of the \sfct{} \fs{} in the whole crossover regime (dotted lines for $\hs[2]/\hs[1]\in\set{-0.025,-0.02,-0.015}$).} 
\label{fig:spnorm_crossover_smallh}
\end{figure}
\begin{figure}
\includegraphics {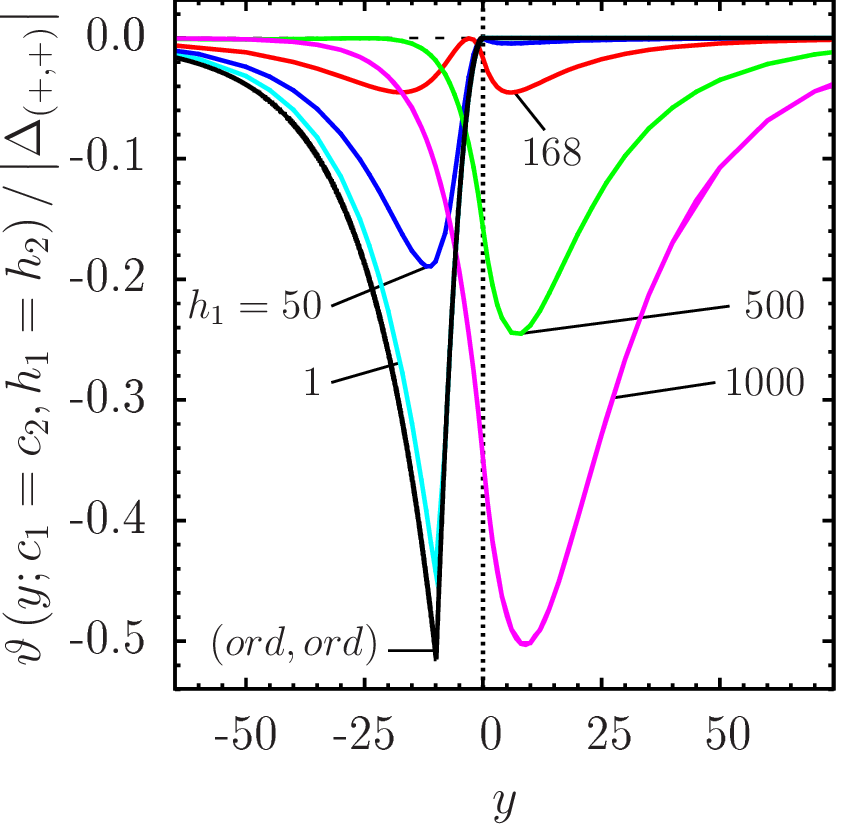}
\caption[ordnorm symm]
	{The \sfct{} \fs{} of the \ccf{} for fixed $\cs[1]=\cs[2]=100$ and increasing surface fields $\hs[1]=\hs[2]$ crosses over from the behavior typical for the \of{ord,ord} BCs \of{\cs[1]=\cs[2]=\infty, \hs[1]=\hs[2]=0} to the behavior of the \of{+,+} BCs. Thereby the minimum below \tcb{} disappears and a minimum above \tcb{} is formed. For moderate values of \hs[1] there are two minima (see, e.g., $\hs[1]=168$). For comparison the \sfct{} $\fs[\of{ord,ord}]\of{y}$ (black line) is also shown. $\fs[\of{+,+}]\of{y}$ is off the scale of the figure ($\fs[\of{+,+}]\of{y=0}/\ccapp=1$). Notice that, even for $\hs[1]=1000$, $\fs\of{y;\cs[1]=\cs[2]=100,\hs[1]=\hs[2]=1000}$ differs markedly from \fs[\of{+,+}]. This is probably due to the strong surface enhancement parameters $\cs[i]$ and the slow convergence apparent in Fig.~\ref{fig:spnorm_crossover}}
\label{fig:ccf_ordnorm_symm}
\end{figure}

\begin{figure}
\includegraphics{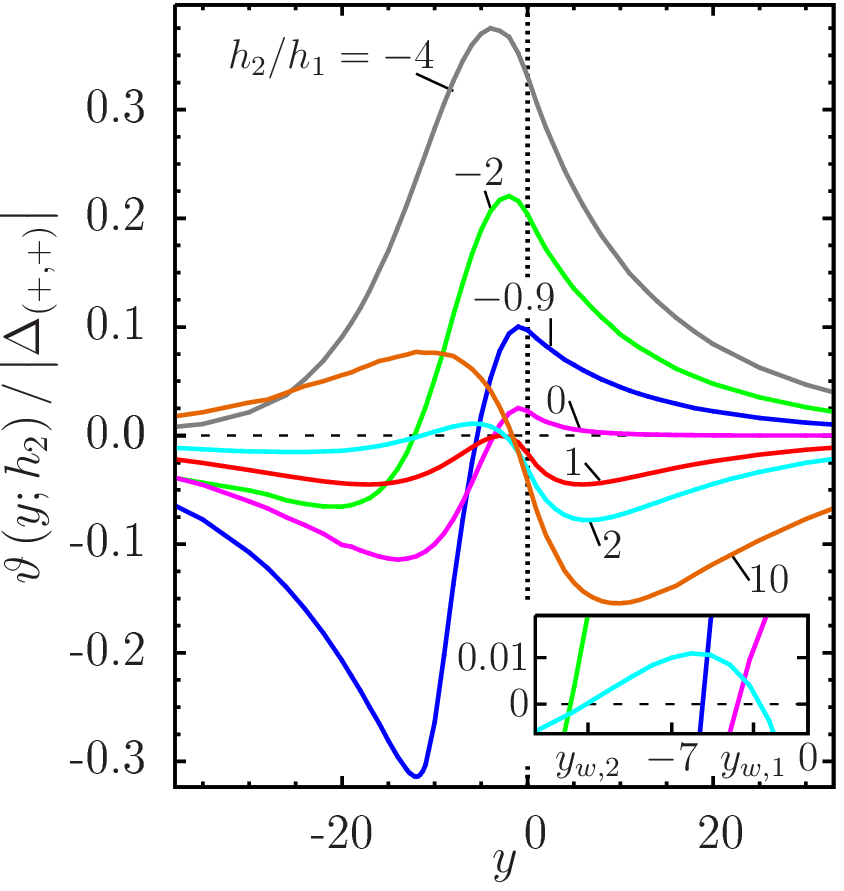}
\caption[ordnorm hr varies]
	{For fixed surface parameters ($\cs[1]=100$, $\hs[1]=168$) of one surface varying the surface field \hs[2] of the second surface with fixed $\cs[2]=\cs[1]$ induces a rich variation of the \sfct{} \fs{} of the \ccf{}. Whereas for $\hs[2]/\hs[1]\lesssim -4$ the characteristics of \fs{} resemble those of \fs[\of{+,-}], for $\hs[2]/\hs[1]=10$ the \sfct{} \fs{} exhibits the characteristics of \fs[\of{+,+}] and \fs[\of{+,ord}] above and below \tcb, respectively. For $-3<\hs[2]/\hs[2]<5$, \fs{} exhibits an even more complicated structure (see the main text, Subsec.~\ref{ssc:ordnorm}). The range of values for $y$ around those corresponding to the wetting transition temperatures, $y_{w,1}=-2.8$ for surface $1$ and $y_{w,2}=-11.3$ for surface $2$ with $\abs{\hs[2]/\hs[1]}=2$, is shown enlarged in the inset for the curves belonging to $\hs[2]/\hs[1]\in\set{-2,-0.9,0,2}$. As discussed in detail in the main text (Subsecs.~\ref{ssc:ordnorm} and \ref{ssc:signchange}), the change of sign occurs up to a finite-size correction $\delta y$ at the wetting transition temperature $y_w$ if the film crosses at this temperature from a UC with an attractive force to a UC with a repulsive force (or the other way round). For $\hs[2]/\hs[1]\in\set{-2,0,2}$ the correction is rather small, $\delta y<1$, but $\delta y=2.6$ for $\hs[2]/\hs[1]=-0.9$.}
\label{fig:ccf_ordnorm_hr}
\end{figure}
\begin{figure}
\includegraphics{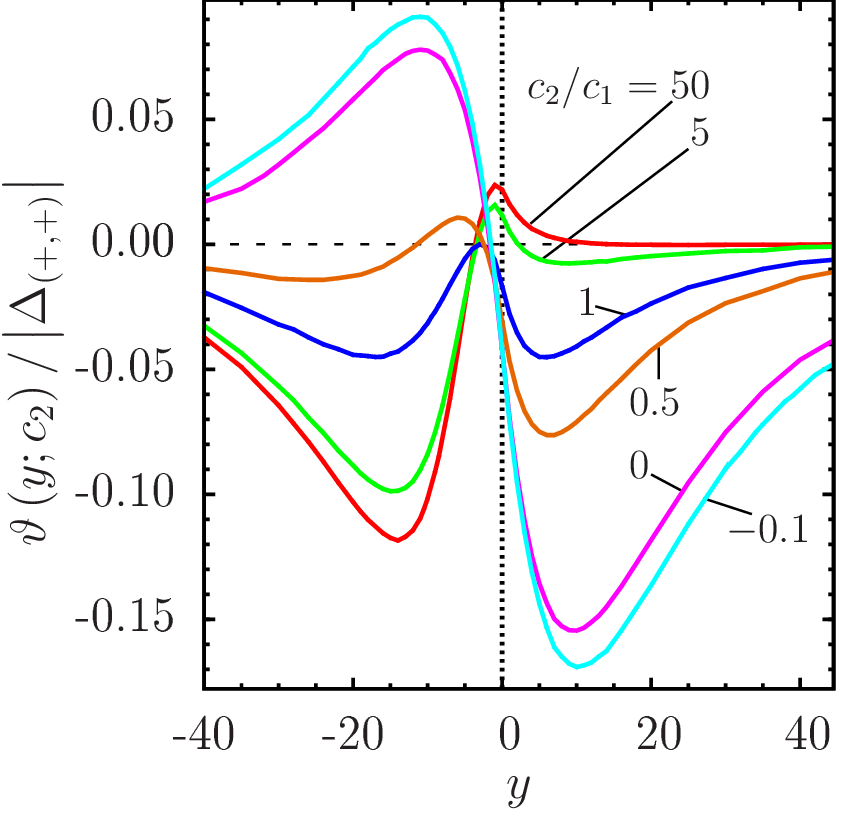}
\caption[ordnorm cr varies]
	{The \sfct{} \fs{} of the \ccf{} for the case that one surface is fixed at $\of{\cs[1]=100,\hs[1]=168}$ and at the other surface the same surface field is applied, $\hs[2]=\hs[1]$, but the surface enhancement $\cs[2]$ is varied. The structure in the variation of $\fs\of{y}$ can be understood by assuming that the film crosses asymptotic regimes of different fixed points upon varying the scaling variable $y$ given in \eref{eq:y} (see the main text). As discussed in the main text (Subsec.~\ref{ssc:ordnorm}), the behavior is similar to the one shown in Fig.~\ref{fig:ccf_ordnorm_hr}. The case $\cs[2]/\cs[1]=5$ shows that it is possible that above \tcb{} the \ccf{} is attractive and that upon lowering the temperature it turns repulsive above \tcb{}.}
\label{fig:ccf_ordnorm_cr}
\end{figure}
\begin{figure}
\includegraphics{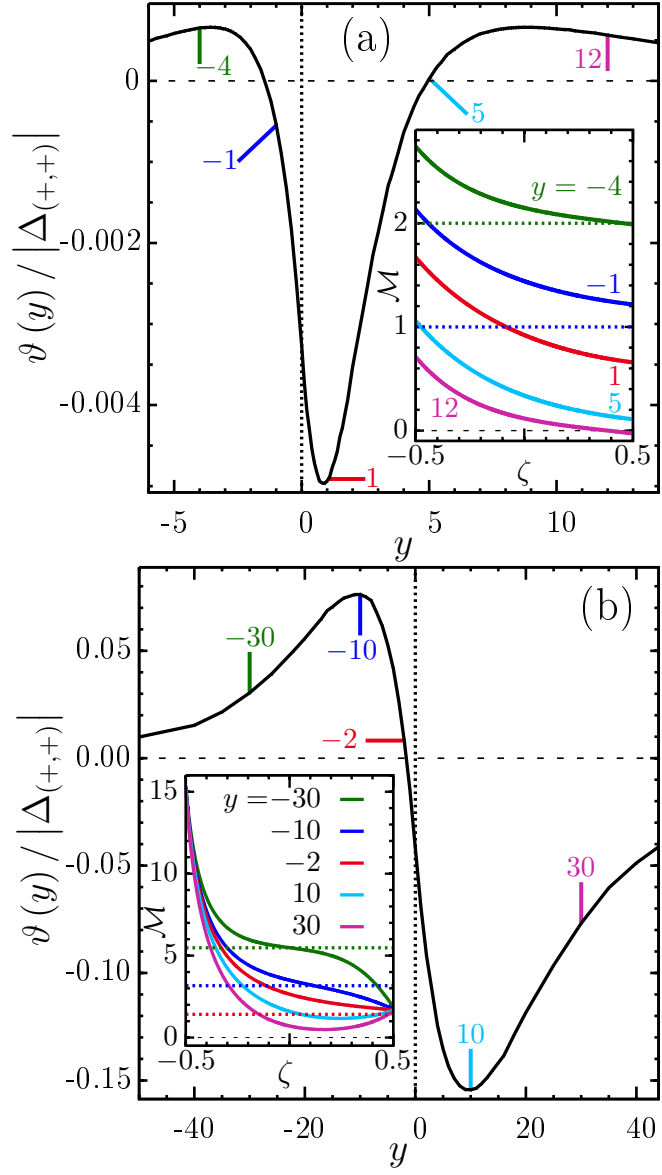}
\caption[sign change]
	{Analysis of the changes of sign of the \sfct{} \fs{} of the \ccf{} as a function of the scaling variable $y=\tau L^{2}$ (\eref{eq:y}) for two examples. 
	Figure~\ref{fig:signchange}(a) shows a typical example for the crossover regime between the normal and	the special transition. The values of the surface parameters are $\cs[1]=\cs[2]=0$ and $\hs[1]=-10\hs[2]=2.5$.
	Figure~\ref{fig:signchange}(b) represents the crossover regime between the normal and the ordinary
	transition. Here the values of the surface parameters are $\cs[1]=\cs[2]=100$ and $\hs[1]=10\hs[2]=1680$.
	In the insets the \sfcts{} $\Ops\of{\zs,y}$ of the \opp{} in the film are shown for those values of $y$ indicated in the main figures. The corresponding bulk values $\Ops[b]\of{y}$ (\eref{eq:scaledopb}), which are $0$ for $y\geq0$, are indicated as horizontal dotted lines. One can infer from these plots that \fs{} changes sign approximately at that value of $y$ at which the value of the \op{} $\Ops\of{\zs,y}$ at one surface equals its bulk value $\Ops[b]\of{y}$ (compare \eref{eq:signccf_and_op}).
}
\label{fig:signchange}
\end{figure}
\begin{figure}
\includegraphics{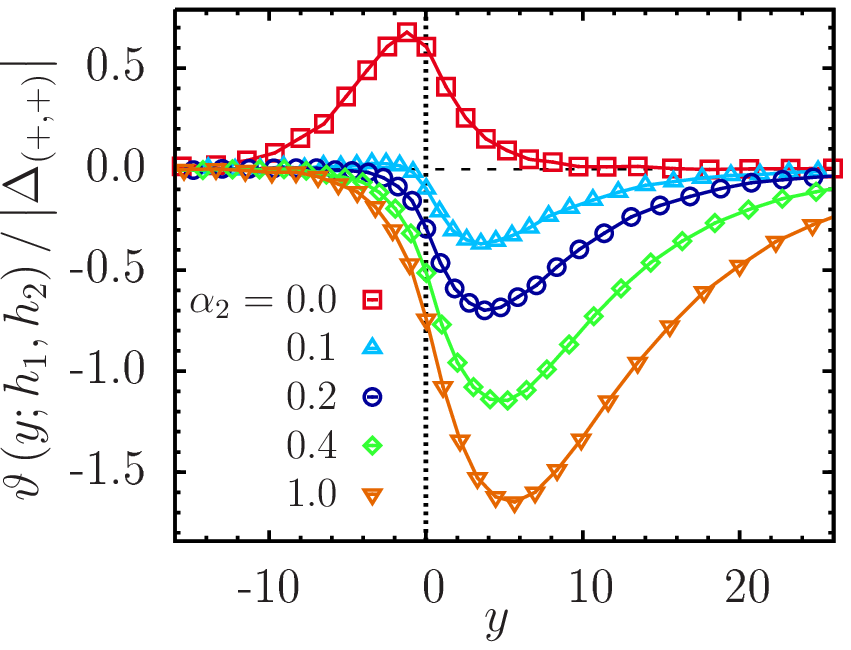}
\caption[dimensional comparison]
	{The \sfct{} \fs{} of the \ccf{} obtained from Monte Carlo simulations \cite{Vasilyev-et:2010} for a $d=3$ Ising film of size $L_x\times L_y\times L_z$, with $L_x/3=L_y/3=L_z=10$ and with periodic BCs in the $x$ and $y$ directions. The surface field \hs[1] of surface $1$ is fixed and corresponds to $\alpha_1=1$ (see Subsec.~\ref{ssc:3d}) and the surface field \hs[2] of the surface $2$, corresponding to $\alpha_2$, is varied. Upon decreasing $\hs[2]$ the minimum becomes shallower and a maximum emerges below \tcb{} (see $\alpha_2=0$; for $\alpha_2=0.1$ there exists a very weak and broad maximum which is hardly seen on the scale of the figure). This behavior is also observed for $d=4$ with surfaces in the crossover regime between the ordinary and the normal transition (see Fig.~\ref{fig:ccf_ordnorm_hr} and the main text). Here $y=t\of{L/\xi_0^+}^{1/\nu}$ with $L=L_z$, $t=\of{T-\tcb}/\tcb$, $\xi_0^+\simeq0.50$ \cite{Ruge-et:1994}, $\nu\simeq0.63$ \cite{Pelissetto-et:2002}, and $\Delta_{\of{+,+}}\of{d=3}\simeqâ0.38$ \cite{Vasilyev-et:2009}.}
\label{fig:force_3d}
\end{figure}
%


\begin{thebibliography}{}
\bibitem{note_rescl_genExponents} By deriving the formulas for the effective rescaling via following the idea for the unscaled \opp[s] one obtains the relation $r=L/L^*$ and one can express the mapping in terms of the general scaling variables: $y^*=r^{-1/\nu}y$, $\Ops\of{\zeta;y,\hs[1],\hs[2]}=r^{\beta/\nu}\Ops[\of{+,\pm}]\of{(r(\zeta-\zeta_0);y^*)}$, and $\fs\of{y,\hs[1],\hs[2]}=r^{d}\fs[\of{+,\pm}]\of{y^*}$; the latter result is obtained by avoiding the use of the stress tensor. However, it is unclear to which extent this concept of rescaling works beyond MFT. It appears to be promising to check whether the results for the $d=2$ Ising film exhibit these rescaling properties, too. If this is the case there is reason to expect that these properties also hold in $d=3$ spatial dimension, where it would in addition provide a tool for analyzing experimental data.
\bibitem{note_h2min} Although \hs[2,min] is an upper bound for the values of \hs[2] for which the rescaling scheme applies, we call it 'minimal' because the physical interest rests in the strength \abs{\hs[2]} of the surface field and \hs[2,min] is the  weakest one for which the rescaling is applicable.
\bibitem{note_dropped_0} In the summary, in favor of an easier reading, we have dropped the index $0$ indicating the unscaled surface field \h[1] as it is present in \eref{eq:fesmallh}.
\end{thebibliography}
\end{document}